# The Mesoscopic Foundations of Non-Equilibrium Thermodynamics

# and The Time's Arrow

# in the Dual Model of Liquids.

***Fabio Peluso***

***WASS Submarine Systems s.r.l. – A Fincantieri Company***
Via Monterusciello 75, 80078 Pozzuoli (NA) – Italy. mailto: fp65.space@gmail.com.

## Abstract

This manuscript has two goals. The first is to show that the elementary interaction in the Dual Model of Liquids (DML) between the solid-like molecule's aggregates and the lattice excitations, is appropriate to represent the link between the behaviour at macroscopic scale of normal liquids, and the physical processes characterizing those systems at mesoscopic scale. On the one hand, at macroscopic scale transport properties of liquids as well as dissipative processes, are described from the statistical point of view by the equations of the Non-Equilibrium Thermodynamics (NET). On the other hand, at mesoscopic scale energy and momentum are supposed in the DML be transported in liquids by means of quanta of elastic energy, the *lattice particles*, and exchanged with solid-like ephemeral aggregates of molecules, the *liquid particles*. This interaction is temporally reversible, as requested by the Onsager' principles and by the Newton' laws of dynamics, and allows for the transport of energy, momentum and mass through a liquid. In the DML framework, the physical processes responsible at macroscopic scale of such transport phenomena were modelled, and the expressions not only for equilibrium properties, such as the specific heat, the thermal conductivity, the diffusion, etc, but also for dissipative processes, such as the thermodiffusion coefficients, the Soret coefficient, the shear viscosity, etc, were deduced and validated by comparison with the experimental data. The elementary interaction is further characterized by a (classical) tunnelling, whose effect is that the energy and momentum exchanged between the two reservoirs in one place, are given back a step further and a time lapse later. The tunnelling is a finger-print of the DML and a missing element in the classical approach; it allowed to correctly modelling, by means of the hyperbolic propagative equations, the time-dependent behaviour of systems following the application of a disturbance. The second goal consists in showing that the duality allows to identify a time's arrow on the mesoscopic scale in liquids. The interaction of quanta of elastic energy with the molecular clusters introduces a privileged direction, a mesoscopic asymmetry, which is relevant in time-dependent and dissipative macroscopic processes, although the elementary interaction remains temporally reversible.

## 1. Introduction.

Equilibrium statistical mechanics is based on two postulates; the first establishes that every macroscopically observable state is actually the cumulative feature of an enormous number of microscopic processes, continuously and spontaneously occurring in the system under investigation, even when its macroscopic state does not change over time (statistical average). The second postulate establishes the *a priori* equiprobability of every microscopic state. From these two principles, a series of theorems arise constituting the basis of the equilibrium thermodynamics. When systems are out of equilibrium, a further postulate is mandatory to ensure that the laws of physics are valid also on microscopic scale. Non-Equilibrium Statistical Mechanics (or of Irreversible Processes) is based on the further principle of "time reversibility", closely resembling that of Newtonian mechanics; it establishes that all the laws of physics must remain unchanged if time $t$ is everywhere replaced by $-t$ and the magnetic field $H$ by $-H$. In contrast to that, in thermodynamics and statistical mechanics, the flow of time has a well-established direction in macroscopic systems. Let's think, for example, of the second law of thermodynamics in the Boltzmann formulation of the principle of increase of entropy, according to which, at the end of a spontaneous process "the most likely situation" is by far that in which the entropy of a closed system is increased, or at most has remained unchanged in the ideal case that only reversible (i.e. ideal) transformations occurred within the closed system (and without considering the negligible local fluctuations, limited in space and time). This rigourus formulation in turn derives from a much less abstract principle, which we all deal with in daily life: any people, although ignoring the laws of physics, will declare on the basis of his own belief that heat flows spontaneously from a hot body to a cold one, and never vice-versa, until thermal equilibrium is reached [1-2].

Phenomena in which transport processes take place are those typically described by the NET: the heat flux through a medium, the mass flux due to a concentration gradient, the volume flux due to a pressure gradient, the momentum transport in liquids under shear geometry, or also the electrical current due to a voltage difference applied to an electrical load. In the approach due to Onsager [3-4], by far the most largely adopted to describe such processes in NET, it is supposed that the gradients of some physical quantities, such as temperature, mass, pressure, shear velocity, voltage, etc, behave as driving forces (sometimes named affinities), $F_i = -\nabla\Phi_i$ producing associated fluxes, $J_i$, such as the flux of heat, mass, volume, momentum, electrical charge, etc, similarly to the "cause-effect" relationship of classical mechanics. This approach uses a mathematical matrix formalism and is based on four main hypotheses [5-8], namely:

i. The quasi-equilibrium postulate, requiring that gradients (driving forces) be not too large.

ii. The linearity postulate, stating that all fluxes are linear functions of the relevant gradients (linear Markoff process); this is a consequence of the previous postulate, allowing a linear dependence of fluxes upon driving forces, due to the smallness of the last,

1. $J_i = \sum_j L_{ij} F_j$ .

iii. The Curie's postulate, constraining the tensor rank of coupled fluxes and forces.

iv. Onsager's reciprocal relations, requiring symmetric coefficient' matrices in the force-flux relationships (this assumption is a consequence of the time symmetry characterizing the processes taking place at microscopic scale in systems described by NET),

2. $L_{ij} = L_{ji}$ .

H.B. Callen and T. Welton [9] introduced the "fluctuation-dissipation" theorem, expressing a functional relation between the irreversible response and the equilibrium fluctuations. Other theorems relating the fluctuations characterizing non-linear irreversible processes have been developed by Bernard, Callen and Lax [9-11].

Many other phenomena in NET are called "crossed", because a driving force is responsible also for other fluxes other than that strictly related to the applied force. Equation (1) expresses exactly this concept, being the generalized flux $J_i$ given by a linear combination of the several generalized forces, $F_j$. Typical examples are the Soret, the Dufour, the Peltier, the Seebeck, the Thomson effects, etc. Very interestingly, another "unexpected" effect, because neither observed nor foreseen before, was recently revealed. It consists in the onset of a temperature gradient in liquids subjected to shear flow [12-16]. The DML has provided the first-ever physical interpretation of this phenomenon: it is nothing else than the manifestation of another crossed effect, namely the conversion of mechanical into thermal energy (other than the momentum flux) in pure liquids under a shear velocity gradient; the theoretical interpretation was verified by comparison with the experimental data available [17-18]. It is worth noticing that this crossed effect was never taken into account before in theoretical modelling, thus explaining why it was originally believed to be unexpected; neither it was detected before the experiments of Noirez and co-workers cited above, because of its extreme weakness, nor actually ever looked for. Like other forgotten effects, it is neither predicted, nor expected in classical models. However, DML allowed for the first time to correctly frame this effect into those typical of NET [see also paragraph 2.5].

The formal and rigorous framework of NET, however, has an intrinsic weak point, normally not highlighted in textbooks. It consists in the lacking of a physical bridging mechanism between the macroscopic behaviour of systems, suitably described by the statistical laws of NET on one side, and the microscopic world of the molecular interactions on the other, obeying the Newton' physical laws of dynamics. This deficiency is precisely the true origin of the Loschmidt' paradox [19], or of the Hilbert sixth theorem [20]. In fact, NET is a phenomenological approach describing (mainly) fluid systems out of equilibrium by means of macroscopic statistical variables, both extensive (volume, mass, energy, etc.) and intensive (temperature, pressure, etc.). Therefore, it does not provide any physical interpretation of the phenomena under study. To fill this gap, many but desultory attempts have been made in the past to find a physical microscopic interpretation of the several phenomena known on a macroscopic scale and described by NET. However, they all foresaw not only involving complex mechanisms at the basis of the observed phenomenology, but also provided in many cases contradictory results within the approach itself (examples are for instance the attempts to model the shear viscosity in liquids [see for instance Section 2 in 17 and related references] or the Soret effect [see Section 2 in 18 and related references], to cite few). It is just in this framework that the DML fits in, with its dualism and the special features of the elementary interaction between *liquid particles* and *lattice particles* [21-22, see also Section 2.1]. I will show that it is precisely this elementary interaction which plays the role of the missing link between the macroscopic and the microscopic worlds; in particular, its fingerprint, the tunnelling, plays a crucial role in the time-dependent states. Additionally, the dynamics beyond such Dual Model provides the key to overcome the well-known time's arrow paradox raised by many authors since the time of Boltzmann. We proceed now to shortly recall the DML model, leaving the reader interested to deepen this specific topic to consult the related literature [17-18, 21-24].

## 2. The *Lattice Particle – Liquid Particle* elementary interaction of the DML as bridge between the mesoscopic world of liquids and the thermodynamic macroscopic approach.

Liquid state is still today one among the more controversial pages of the physics. The liquid state is by far that existing in the more narrow temperature range, compared with the others states of matter. All the physical parameters characterizing the liquid state show values and their variation with temperature and pressure much similar to the analogous of solids, although liquids flow and do not have a own shape, like gases. For long time liquids were believed condensed gases, however, since the 40's of the past century, this idea left progressively way to the opposite idea of liquid being molten solids. Recently, Chen [25] critically reviewed the various approaches to the liquid modelling developed over the past 150 years. The oldest of them tries to understand transport in liquids by modifying the Boltzmann theory of gas, as Enskog did (see ref. [9] in [25]). Einstein was the pioneer of another attempt, ([see ref. [10] in [25]) consisting in determining the molecular size by studying the Brownian motion. Finally, the third line of thought is that consisting in starting from the solid picture; it can be dated back to Maxwell [26], successively reworked by Debye [27-28], Brilluoin [29-30] and Frenkel [31]. A very interesting approach in this frame is that due to Zwanzig, who explicitely analyzed the consequences of considering the presence of collective excitations in liquids similar to those of solids [32]. The current idea about the dualism characterizing the liquid state has made its way in the scientific community; it allows liquids to share with solids the modes of transmitting and storing the energy [32-45] by means of collective excitations. The DML has further pushed such crazy idea by introducing other concepts: i) the duality of liquids consists in the presence of solid-like ephemeral aggregates of molecules, (those we like to dub icebergs) whose number and size progressively decrease with temperature; besides, ii) energy and momentum are exchanged among such icebergs by means of collective oscillations; iii) such modes are (quasi) elastic inside the icebergs, while they become inelastic when leaving or hitting them. These concepts are resumed and physically described by the elementary interaction between *liquid particles* and *lattice particles*, shown in Figure 1, that we are going to discuss and show their pivotal role in the mesoscopic dynamics of liquids.

The need for a physical model of liquids arises not only from the search for a model as an end in itself, but also with the aim of providing the physical interpretations of transport phenomena in liquids, in agreement with the experimental data. Transport phenomena in physics are of huge importance, because they impact on all the processes leading to a physical change in the universe. For this reason, they are considered among the fundamental building blocks governing the evolution of the universe, as well as responsible for the success of life on earth. Transport phenomena are statistical irreversible processes due to the random continuous motion of molecules, typical of fluids. They are governed by two fundamental concepts: the conservation laws, stating that the total sum of the quantities under study must be conserved by the complete system and its environment, and the constitutive equations, describing how the quantities in question respond to external stimuli. These two set of equations demonstrate not only the connection of transport phenomena with thermodynamics, but also why transport phenomena are irreversible. The physical systems in which transport phenomena occur, ultimately look for their lowest energy state, following the minimum energy principle. As they approach this state, they also approach the thermodynamic equilibrium, so that driving forces stop in the system, and consequently transport ceases.

This section is aimed at framing the Dual Model of Liquids and the elementary interaction on which it is based, in the widest picture of the Non-Equilibrium Thermodynamics. One the one side, we will show how the elementary interaction represents the link between the mesoscopic scale of liquid modelling and the macroscopic scale of the statistical physics. On the other hand, we will show how the DML provides a physical interpretation of the transport phenomena typical of liquids, phenomenologically interpreted by the NET.

## *2.1 Introduction to the Dual Model of Liquids.*

Compared to the situation with gases and solids, our current understanding of transport properties in liquids remains rather limited. One of the reasons is the absence of small parameters, as pointed out by Landau [46]. In fact, on one side, differently from gases, intermolecular forces in liquids cannot be approximated as contact or short-ranged, because they are long ranged similarly as in solids. On the other side, differently from solids, displacements of molecules from their equilibrium positions involve long distances, being the molecules not fixed in their positions as in solids, and therefore cannot be approximated alike small oscillations. The peculiarity of liquid dynamics, where the molecules exhibit fast solid-like oscillations about temporary equilibrium positions, in addition to slow diffusion of these equilibrium positions, plays the essential role. Because a reliable and complete theory of liquid structure is not developed yet and it is very unlikely that it will be completely developed in a near future, new ideas and proposals should be welcome. If today a definite and general view of the mesoscopic structure of liquids is provided by the experimental data, as it will be shown in the following, from what reported above it should not be surprising that at the beginning of the past century, famous theorists such as Frenkel [31], Brillouin [30] and Eckart [47-48], identified the length scale that is today called "mesoscopic" as the most promising to be investigated in liquids. In this intermediate scale, where liquids deviate from their macroscopic behaviour, although the dynamics still involve a large number of molecules, the relaxation mechanisms become fundamental in the dynamics of molecules in liquids and in the processes of redistribution of energy and momentum among the available DoF. Such a length scale is that where the statistical mechanics is the most appropriate tool for describing the motion of molecules "in the average". This scale is intermediate between the atomic scale, where quantum mechanics dictates the rules of game and the motion of particles is described by the Newton's laws, and the macroscopic scale of liquids, described by fluid-dynamics and thermodynamics. Eckart was the first to explicitly assert that "the distinction between liquids and solids is quantitative and not qualitative" in his papers on the anelastic fluid [47-48], the border line being the relaxation time $\tau$ , or the frequency $\nu = \frac{1}{\tau}$, whose central role in the dynamics of liquids was recognized for the first time by Maxwell [26]. Going back even further in time, Brillouin early in 1936 [30] advanced the ingenious idea that liquids could be arranged in "islands" of solids: "*Microscopiquement, (le liquids) sont des solides; macroscopiquement, ils appairetront liquides*". Trachenko [49] raised an interesting connection between such characteristic of the liquids and the property of phonons travelling through them: unlike in solids, the phonons in liquids have the relevant property that their phase space is not fixed but variable; in particular, the phase space reduces with temperature. This could be the explanation of the peculiar behaviour of the specific heat in liquids, which also decreases with temperature. In the DML, such behaviour is quantified and expressed by the trend of the parameter $m$ vs $T$, $dm/dT < 0$. The above reasoning correlates $m$ with the vibration density of states, VDOS [50-51], recently discussed by Baggioli and Zaccone by means of the instantaneous normal modes, INM [52-53]. Notably, the INM in liquids were early considered by Stratt already in 1996 [54]. Landau [55] applied the concept of dualism of liquids to the liquid Helium. Many other scientists dedicated their studies to the physics of mesoscopic scale, just to cite few, Eyring [56-57], Fiks [58-59], Andreev [60], Nettleton [61-63], up to the more recent Phonon Theory of Liquid Thermodynamics, PLT [35-45]. Despite the fact that the input from theorists arrived so early, liquids are being investigated experimentally on such scale only since the end of the past century. When investigated on such a scale, therefore at high frequencies, liquids respond to perturbations with a visco-elastic behaviour. Thousands among experimental papers as well as theoretical developments and books have been published since the pioneering works of Maxwell, Frenkel, Brillouin and Eckart. The first experimental findings in scattering experiments and their theoretical interpretations are dated back to the mid of the past century [64-87]; Chen [25] has given an interesting critical review of these works. In the recent experimental campaign, [88] the "*presence and long lasting permanence of elasticity on mesoscopic scale*" in water and glycerol, was revealed for the first time. The elastic response in such liquids persisted indeed for time laps of microseconds, hence four orders of magnitude longer than the typical intermolecular relaxation time. Such rubber-like elasticity and large strain response in fluid glycerol were interpreted by the authors as due to the *relaxation processes of collective modes in metastable groups of molecules* (that may be identified with those we have dubbed icebergs in the DML). Consequently, one may argue that the lifetime of such icebergs is equal to, if not larger than, the relaxation time of elastic propagation, as already hypothesized in [22]. The authors also point out that such modes require the existence of a transient state with *solid-like long-range correlations*, although different from the bulk state. This study has revealed that liquid shear elasticity could be very underestimated on such mesoscopic scale and high frequency modes domain with respect to what measured with the classical methods on macroscopic scale.

Jumping back by 50 years, Rahman and Stillinger [89] set-up the first numerical model to calculate the dynamical structure factor $S_{EQ}(q,\omega)$ for liquid water based on the Molecular Dynamics (MD) approach. Their model analysed the fluctuation phenomena occurring in a system of 216 water molecules kept at ambient steady conditions. Two interesting results were obtained, namely that "transverse currents" were present in the liquid in the form of "propagating collective modes", and that the spectrum of density of fluctuations exhibited a secondary maximum at a "much higher frequency" than the usual sound propagation frequency. The propagation velocity of this transverse currents were found by MD simulations amounting to about $\approx 3200 m/s$, surprisingly close either to that of elastic waves in solid water and more than double than that in liquid water. This value was experimentally confirmed by Ruocco et al. [69] in 1996 by means of X-ray scattering technique. Proctor [90], has constructed a numerical method to calculate, within the frame of the PLT Dual Model, the internal energy and heat capacity of liquids over wide pressure and temperature ranges. Proctor has applied his numerical code to a number of real liquids in both the subcritical and the supercritical regime, in which the Frenkel line represents the border for the liquid state, in agreement with the PLT model. Experimental data have been fitted allowing to test the theoretical model with unprecedented rigor.

To stay again in the field of numerical simulations, Ghandili et al. [91] developed a model to simulate a system arranged by means of *molecular clusters*, whose molecules may experience only quasi-harmonic vibrations around equilibrium position within the lattice, exactly as supposed in the DML, and free particles. The authors approach the substance as a fractal lattice and introduce the new concept of thermodynamic dimension, $D_T$, corresponding to the average number of intermolecular interactions per liquid particle as function of temperature; it turns out to be a measure of how many molecules are bound in the lattice of a single *liquid particle*. One of the first results is that $D_T$ allows to generalize Einstein and Debye models, so that both are a special case of the numerical model by Ghandili. Then the authors propose a general expression for the vibrational DoF contribution to the isochoric specific heat. They finally show that in a wide range of temperature and pressure, the calculated results agree very well with experimental data for isochoric heat capacities in dense region in the considered fluids. In particular, by introducing the condition $D_T = 1/2$ to plot the Frenkel line, they predict solid-like features around the critical point. Within the same framework, Ghandili approached also the viscosity modelling in liquids [92] and the estimation of the number of particles in temporary clusters, although evaluated in phase transitions [93]. Zhao et al [94] use the expression for the specific heat of liquids deduced in the PLT dual model to obtain an expression for the thermal conductivity of liquids obtained within the dual framework. The agreement with experimental data is the highest ever obtained, even if compared with previous models of liquids.

We may therefore state with high confidence that modelling liquids as Dual systems, composed by metastable solid-like aggregates mixed with molecules arranged along with the classical liquid, as supposed in the DML, is a widely established practice for many years. Supporting this statement there are today some facts firmly assessed on the basis of experimental outcomes. One is that heat can be described in liquids as mechanical vibrations similarly to sound, propagating its effect through the atomic lattice [95 and references thererin]. The second consists in the fact that liquids exhibit long-ranged spatial correlations on mesoscopic scale, on distances of the order of some molecular diameters, that were early theoretically predicted and recently experimentally verified. In fact, the phononic branch of heat differs from that of sound because heat waves oscillate most at frequencies around GHz, while those of sound waves oscillate around much lower frequencies, typically in the KHz frequency band (Figure 2) [95]. Therefore, heat perturbations propagate over distances much lower than those of sound (of the same OoM as the ratio between the two frequencies bands, [24]). In the DML, elastic perturbations, and with them even the heat, propagate similarly as in solids within the wandering icebergs, giving way to the classical propagation by means of longitudinal wave-packets in between them, as shown in pictorial way in Figure 3. These facts allow also explaining why both acoustical and optical modes are present and have been detected in liquids, the *liquid particle* borders working as converters between such two branches. The solid-like structures, that we like to dub icebergs, are ephemeral aggregates of molecules, that communicate each other, exchanging energy and momentum, as in solids. Such aggregates have lifetimes of the order of some relaxation time intervals, and molecules may fluctuate from one cluster to the nearest neighbour, [22,30,88]. More specifically, the duality of the DML consists in accounting for, and distinguishing between the *liquid particles*, i.e. the solid-like aggregates of molecules, and the amorphous phase, typical of liquids, where molecules fluctuate independently from each other. In the same way, elastic perturbations propagate across the *liquid particles* by means of quasi-harmonic quantized waves, the *lattice particles*. These are in turn distinguished from the wave-packets, i.e. the quasi-particles, anharmonic wave-packets in which *lattice particles* commute when they interact with the *liquid particle* borders, as shown in Figure 3. Therefore, the perturbations travelling inside a solid-like structure show a solid-like character, i.e. they are (quasi) harmonic waves whose speed is close to that of the corresponding solid phase (3200 m/s in the case of water, [69-70,88]). Because they lose the harmonic outfit and generate wave-packets, this velocity reduces to the classical velocity of elastic waves across liquids when they leave the *liquid particle* and travel through the amorphous environment. The *liquid particles* are considered as solid-like aggregates because their internal structure should be considered as "frozen" for time intervals whose length is comparable to the relaxation times characterizing the liquids, thus allowing the propagation of (quasi)-harmonic waves through them. Two reservoirs should be considereed for the internal energy of a liquid, one related to harmonic waves, the other to non-harmonic contributions [22,24]. The same approach was followed by Bolmatov and Trachenko when they introduced the Phonon Theory of Liquid Thermodynamics [35-45]. A local configuration lasts until a fluctuation in the kinetic energy induces a rearrangement of the positions of some of the atoms towards a new local minimum in the potential energy surface. These fluctuations are identified in the DML as the local microscopic heat currents, $\left\langle \frac{\delta T}{\delta z} \right\rangle$, in form of wave-packets of elastic (thermal) energy [22].

Momentum and energy are exchanged between the two populations of pseudo-particles, the *liquid particles* and the *lattice particles*, thanks to the anharmonicity which characterizes the interaction between these waves and the solid-like icebergs. The interaction, through which a fraction of the energy, mass and momentum are transported, is that schematically represented in Figure 1. The mean free-flight of a wave-packet between two successive interactions with as many *liquid particles*, is long $\langle \Lambda_{wp} \rangle$ and lasts $\langle \tau_{wp} \rangle$. Along $\langle \Lambda_{wp} \rangle$ they are considered as local microscopic heat currents. $\langle \nu_{wp} \rangle = 1 / \langle \tau_{wp} \rangle$ is the average frequency of the elementary interactions represented in Figure 1. If S is a surface arbitrarily oriented within the liquid, at thermal equilibrium an equal number of wave-packets will flow through it in opposite directions. Such heat currents are driven locally by the virtual temperature gradient $\left\langle \frac{\delta T}{\delta z} \right\rangle \cong \frac{\Delta T}{\langle \Lambda_{wp} \rangle}$ applied over $\langle \Lambda_{wp} \rangle$, ensuring the isotropy of their distribution in the liquid isotropic.

Starting from the experimental outcomes, DML faces the problem of liquid modelling using a "bottom-up" approach, the building blocks being the aggregates and their interactions with the lattice collective excitations. The elementary interaction has also the tunnelling, Figure 4, as a peculiar characteristic. These facts make the DML the first and so far the unique theoretical approach that, by analyzing the energy and the momentum exchanged between *lattice particles* and *liquid particles*, provides a physical and quantifiable interpretation of the transport processes in liquids, rather than phenomenological. It allows deducing physical parameters and fundamental features supported by the comparison with the experimental data, such as a mesoscopic model for thermal conductivity and its expression, the specific heat, the propagation of energy, mass and momentum in condensed media, the comparison with systems exhibiting ***k***-gap and non-affine dynamics systems framework, the vibration density of states, VDOS, the phonon mean free path and their life-time. As specified in previous papers, this model does not claim to replace other models of liquid structure, it simply provides a different perspective based on the capability of liquids to propagate energy, momentum and mass by means of collective excitations [17-18,21-24]. Let's therefore analyse how *liquid particles* and *lattice particles* interact.

As is well known [5-8], the total thermal energy content at the temperature $T$ per unit of volume of a substance, $q_T$, is given by:

$$3. \quad q_T = \int_0^T \rho C_V \, d\theta = f\left[\Theta / T\right]$$

where $\rho$ is the medium density, $C_V$ the specific heat at constant volume per unit mass, $\Theta$ the Debye temperature of the liquid at temperature $T$. Since the dynamics described above occurs at high frequencies and involves only the DoF of the lattice [95], we introduce the parameter $m$ which defines the ratio between the number of collective DoF surviving at temperature $T$ and the total number of available collective DoF. Therefore, $m$ accounts for the average number of DoF excited and actually participating to the collective dynamics, and the fraction $q_T^{wp}$ of $q_T$,

$$4. \quad q_T^{wp} = m q_T = \left\langle n^{wp} \cdot \varepsilon^{wp} \right\rangle = m \frac{\int_0^T \rho C_V \, d\theta}{\rho C_V T} \rho C_l T = m^* \rho C_l T$$

accounts for that part of the thermal energy transported by the wave-packets (the definition of $m*$ is trivial). It was shown [22] that it holds $0 \le m \le 1$, as one expects from its definition. $n^{wp}$ is the wave-packets density and $\varepsilon^{wp}$ their average energy. $q_T^{wp}$ is propagated through the liquid by means of inelastic interactions between the *liquid particles* and the *lattice particles*, the last being the carriers of the thermal or, generally, the elastic energy through the liquid. Figure 1 and Figure 4 exemplify the elementary physical mechanism behind the DML. The *lattice particle* ↔ *liquid particle* interaction allows exchanging both energy and momentum between the two populations, and consequently induces also mass diffusion. In the first part of the elementary interaction, the *lattice particle* collides with the *liquid particle* and transfers to it momentum and energy (both kinetic and potential). In the second part, the *liquid particle* relaxes, returning the energy to the thermal reservoir through a *lattice particle*, alike in a tunnel effect. Considering an event of type a) of Figure 1, an energetic wave-packet collides with a *liquid particle* transferring to it the energy $\Delta\varepsilon^{wp}$ and the momentum $\Delta p^{wp}$:

5. $\Delta\varepsilon^{wp} = h\langle\nu_1\rangle - h\langle\nu_2\rangle = \Delta E_p^k + \Delta\Psi_p$
6. $\Delta p^{wp} = f^{th} \cdot \langle\tau_p\rangle$

$\langle\nu_1\rangle$ and $\langle\nu_2\rangle$ are the wave-packet (central) frequency before and after the collision, respectively; $\langle\tau_p\rangle$ is the duration of the interaction during which the inertial force $f^{th}$ is exerted. The energy and momentum lost by the wave-packet after the collision a) are acquired by the cluster. Part of $\Delta\varepsilon^{wp}$ converts into kinetic energy $\Delta E_p^k$ acquired by the *liquid particle*, the remaining part converts into $\Delta\Psi_p$, i.e. the potential energy of internal DoF of the solid-like cluster. The kinetic energy is dissipated as friction. The potential energy $\Delta\Psi_p$ is relaxed by the *liquid particle* once the interaction is completed, so that the residual energy stored into internal DoF returns to the liquid reservoir, although in a more degraded form. The dissipation process lasts $\langle\tau_R\rangle$ during which the *liquid particle* travels by $\langle\Lambda_R\rangle$, its total displacement, $\langle\Lambda\rangle$, and the total duration of the process, $\langle\tau\rangle$, being given by:

7. $\langle\Lambda\rangle = \langle\Lambda_p\rangle + \langle\Lambda_R\rangle$
8. $\langle\tau\rangle = \langle\tau_p\rangle + \langle\tau_R\rangle$

A signature of the DML and of the *liquid particle* ↔ *lattice particle* collision is the dynamics shown in Figure 4; it shows how the tunnel-like effect characterizes the elementary interaction: the energy subtracted from the phonon' reservoir returns to it a time interval $\langle\tau\rangle$ later and a step $\langle\Lambda\rangle$ forward. The dynamics is deeply analysed in [22], where (Table 1) also the OoM of the relaxation times are evaluated. Of course, there are no explicit quantum mechanical effects beyond such tunneling, the terminology has the only scope to point the attention on the fact that during $\langle\tau\rangle$ the energy disappears from the liquid thermal reservoir because it is trapped in the internal DoF. Once $\langle\tau\rangle$ has elapsed, it reappears in a different place. In this way, relaxation times, introduced ad hoc by Maxwell and Frenkel find a physical interpretation in the DML (interestingly, $\langle\Lambda\rangle$ and $\langle\tau\rangle$ in eqs.(7) and (8) have the same meaning as $\delta$ and $\tau_F$ defined by Frenkel, see Figure 3 and [31]). This tunnelling has a strong relevance during the transient phase, such as for instance that following the application of an external stimulus to a system, as deeply discussed in [24]; it shall be dealt with in depth in Section 2.3.

The elementary interaction shown in Figure 1 and Figure 4 depend upon several parameters that are pivotal for the DML; their evaluation, in turn, rests on that of $\langle\tau_0\rangle$, $\langle\tau_p\rangle$ and $\langle\tau_R\rangle$, and hence on the physical modelling of the relaxation processes involving internal DoF. An interesting theory has been proposed by Nettleton [61-63], who introduced a set of phenomenological equations, valid at all frequencies, for the rates of change of internal relaxation parameters describing molecular excitations and structural changes in a liquid. He found an Onsager coupling between the rate equations and the pressure tensor. A it concerns the phonon mean-free-path $\langle \Lambda_0 \rangle$ and life-time $\langle \tau_0 \rangle$ within a *liquid particle*, their values can be deduced from light scattering experiments [67-87]. Of courxe, $\langle \Lambda_0 \rangle$ will be a multiple of the phonon wavelength $\lambda^0$, $\langle \Lambda_0 \rangle = n\lambda^0$, and $\langle \tau_0 \rangle$ of $\tau = 1/\nu^0$, $\langle \tau_0 \rangle = n/\nu^0$, with $n > 1$. Using the data for water of [67,82], typical values for the parameters characterizing a phonon in the DML (variation range is function of temperature, pressure and $q$ orientation) are: (central) frequency $\frac{n}{\langle \tau_0 \rangle} = \langle \nu^0 \rangle \approx 0{,}95 \div 2{,}5\; THz$, wave-length $\frac{\langle \Lambda_0 \rangle}{n} = \langle \lambda^0 \rangle \approx 1 \div 3\; nm$ and velocity $\frac{\langle \Lambda_0 \rangle}{\langle \tau_0 \rangle} = \langle \lambda^0 \rangle \cdot \langle \nu^0 \rangle = \langle u^0 \rangle \approx 3.1 \div 3.4 \cdot 10^3\; m/s$. Very interestingly, this value fits very well the experimental data obtained for the propagation velocity of thermal waves in water [69]. The value of $\langle \Lambda_0 \rangle$ thus obtained represents also the typical OoM of the size of a liquid particle at the exchanged momentum $q$. We may then confidentially assume a value of few units for the parameter $n$.

Several very important aspects of the elementary interaction of Figure 1 should be highlighted. Firstly, the events a) and b) are the time-reversal of each other, as requested by the Newtonian dynamics. Second, in equilibrium conditions and for an isolated system, events a) and b) will alternate over time, to keep statistically equivalent the balance of the two energy reservoirs. Events like those of Figure 1 will be equally probable along any direction, with a null statistical average over time and space. Consequently, the mesoscopic equilibrium induces the macroscopic equilibrium, following the Onsager postulate [3-4] and the Newtonian dynamics. On the contrary, when the symmetry is broken or, generally, the system is no longer isolated, one of the two types of interactions will prevail over the other, depending on the physical conditions occurring in the system. The perturbation will generate an imbalance of the number of collisions and possibly also a variation of the energy of the phonons (we assume that their number does not change as a first approximation for small perturbations) and an imbalance of the number of collisions. We will return on this point when discussing the out-of-equilibrium phenomena. Incidentally, events like those of Figure 1a are responsible for thermo-mechanical effects in liquids, such as for instance the thermodiffusion (see [18] and Section 2.4]), while those of Figure 1b may explain, for example, viscosity in liquids [17] as well as the "unexpected effect" detected by Noirez' group in liquids under shear [18]). In Section 3, I will show how this approach contributes to clarify even the long debated issue of the time's arrow or of the Loschmidt paradox [19-20,96-100].

Let us return on the dynamics of the interaction between the phonons and the wandering solid-like icebergs, Figure 1 and Figure 4. Momentum and energy are exchanged between the two reservoirs, the *liquid particles* and the *lattice particles*, because of the anharmonicity of the interaction. The momentum $\Delta p^{wp}$ generates the force $f^{th}$

$$9. \quad f^{th} = \frac{\Delta p^{wp}}{\langle \tau_p \rangle}$$

i.e. the inertial term responsible for the energy and mass propagation in liquids [22]. One aim of this paper is to show that the elementary interaction between *liquid particles* and *lattice particles* represents the so far missing link, and therefore a reliable candidate, as a bridge between the microscopic molecular world of liquids, in which the Newtonian laws of dynamics hold, and the macroscopic world described with the statistical approach of the NET. The scope is pursued by showing that it becomes possible to provide the physical interpretation of energy, mass and momentum transport in liquids, at least for the fractions carried by the collective excitations, either in stationary, time-independent conditions or in transient phases. A separate section is dedicated to the other aim of the manuscript, i.e. to showing that the duality of liquids in the DML introduces a preferential direction at mesoscopic scale for the time's arrow in time-dependent processes. These processes are those that, at macroscopic scale, are characterized by a time's arrow, apparently violating the reversibility Newtonian principle. It is important to point out once again that the elementary interaction has a time-reversal character, it is in fact a spatial inhomogeneity on microscopic scale in a dual system that is responsible of a directionality of the flow of time.

### *2.2 Energy Propagation: Fourier time-independent diffusion equation.*

What illustrated in this section is applicable to the general situation of the propagation of energy, momentum or even mass through a liquid; however, we will specifically analyse the most familiar case of the propagation of thermal energy. When the only effect on a system due to the external environment is the (thermal) energy flowing through it, Eq. (1) becomes:

$$10. \quad J_q = -L_{qq}\frac{\nabla T}{T^2} = -K\nabla T \ ,$$

where $J_q$ is the heat flux, $K$ the thermal conductivity of the liquid, $L_{qq}$ the phenomenological Onsager coefficient describing the proportionality between the generalized force $\nabla T$, and the fluxes it generates, $J_q$. We adopt the following formalism of eq.(10)

11. $J_q^{wp} = -K^{wp} \nabla_z T$

to account only for the wave-packets contribution to the total amount of heat diffusing through the system. Therefore, alle the physical quantities have the same meaning of the analogous ones defined above, but referred only to the wave-packet current. We have supposed for simplicity that the temperature gradient is applied along $z$, as well as that any effect due to internal friction among molecules, i.e. viscosity, is neglected because the system is supposed at rest and in a stable configuration to avoid convection; in the same way, any effect due to the thermal expansion is not considered, and all variables in Eqs.(10) and (11) are time-independent. Finally, dispersionless systems are considered, and any variation with temperature or pressure of the several quantities are neglected.

Equation (11) is a diffusive-type equation for a time-independent system with a uniform temperature gradient established throughout a homogeneous fluid; therefore, the heat flux is everywhere and every time proportional to the temperature gradient. Equations like (11) are very familiar in NET because they describe the Onsager's theory property that the response of a system to a generalized force is *simultaneous* with its application and proportional to its magnitude (stationary solutions). As a general rule, such *simultaneity* in a macroscopic theory turns out to be the stationary approximation to a *causal* behavior, where the response to a force comes *after* the application of the force.

Combining Eq.(11) with the continuity equation, $\rho C_p^{wp} \frac{\partial T}{\partial t} + \frac{\partial J_q^{wp}}{\partial z} = 0$, with $C_p^{wp}$ being the isobaric heat capacity per unit of mass, we get, for instance, the parabolic propagation equation in terms of temperature $T$:

12. $\frac{\partial^2 T}{\partial z^2} = \frac{\rho C_p^{wp}}{K^{wp}} \frac{\partial T}{\partial t}$

In [22-23] we have discussed in detail eqs. (11) and (12), in particular how the liquid duality affects $K^{wp}$ and $C_p^{wp}$. The wave-packets in liquids are pushed and driven between two successive collisions, such as those of Figure 1, by the virtual temperature gradient $\left\langle \frac{\delta T}{\delta z} \right\rangle$. The adjective "virtual" is not used by chance; in fact, $\left\langle \frac{\delta T}{\delta z} \right\rangle$ is not associated with a real physical gradient, but rather accounts for a sort of curvature in the phase space of the dual system, which influences the trajectory of the phonons [46]. During their free-flight between two successive interactions with the *liquid particle*, the wave-packets are considered as local microscopic energy currents. Each *lattice particle* freely propagates along an average distance $\langle \Lambda_{wp} \rangle$, during a time interval $\langle \tau_{wp} \rangle$, which is its average life-time. Accordingly, the ratio $\langle \Lambda_{wp} \rangle / \langle \tau_{wp} \rangle$ defines the wave-packet average propagation velocity between two successive interactions, $u^{wp} = \dfrac{\langle \Lambda_{wp} \rangle}{\langle \tau_{wp} \rangle} = \lambda^{wp} \cdot \nu^{wp}$, where $\nu^{wp}$ and $\lambda^{wp}$ are the central frequency and wave-length of the wave-packet[1], respectivley. Because at equilibrium an equal number of wave-packets will flow through a surface S arbitrarily oriented within the liquid in opposite directions, from Eq.(4), the variation of the energy density within the liquid along the "+$z$" direction is:

$$13.\ \ \delta_{+z} q_T^{wp} = \frac{1}{6}\delta(mq_T) = \frac{1}{6}\delta\left[n^{wp}\left\langle \varepsilon^{wp} \right\rangle\right] = \frac{1}{6} n^{wp} \left\langle \Delta\varepsilon^{wp} \right\rangle = \frac{1}{6}\frac{\partial q_T^{wp}}{\partial T}\left\langle \frac{\delta T}{\delta z} \right\rangle \left\langle \Lambda_{wp} \right\rangle,$$

where (third to fourth member step) the wave-packets distribution function $n^{wp}$ is supposed not significantly altered along the heat current, only their average energy changes (we will return on this point at the end of this section). The temperature gradient $\left\langle \dfrac{\delta T}{\delta z} \right\rangle$ represents the thermodynamic *"generalized"* force driving the diffusion of thermal excitations along $z$. The factor $1/6$ accounts for the isotropy of the phonon current at equilibrium. The heat current $j^{wp}$ represents the wave-packet diffusion equation, and is obtained by applying this driving force over the distance $\langle \Lambda_{wp} \rangle$ along the "+$z$" direction:

$$14.\ \ j_{+z}^{wp} = -D^{wp}\left(\frac{\delta q_T^{wp}}{\delta z}\right)_{+z} = -\frac{1}{6} u^{wp} \left\langle \Lambda_{wp} \right\rangle \frac{\partial q_T^{wp}}{\partial T}\left\langle \frac{\delta T}{\delta z} \right\rangle.$$

$D^{wp} = u^{wp}\langle \Lambda_{wp} \rangle$ is the diffusion coefficient of the wave packets. The propagation of an elementary heat current lasts the time interval $\langle \tau_{wp} \rangle$ and is driven by the virtual temperature gradient $\left\langle \dfrac{\delta T}{\delta z} \right\rangle$ over the distance $\langle \Lambda_{wp} \rangle$. The heat current may also be obtained by calculating the thermal energy flowing in a cylinder per unit cross section and per unit of time:

$$15.\ \ j_{+z}^{wp} = u^{wp}\delta_{+z} q_T^{wp} = \frac{1}{6} u^{wp} n^{wp} \left\langle \Delta\varepsilon^{wp} \right\rangle.$$

$q_T^{wp}$ in Eq.(15) allows to calculate the phonon specific heat contribution $C_V^{wp}$ [22-23]:

[1] An elastic perturbation of the liquid lattice should be represented by a wave-packet, whose localization depends on the microscopic structure to which it is associated. Here it will be represented, for the sake of mathematical simplification, before and after an interaction, as a monochromatic wave, longitudinally polarized, and hence identified by the average wave-length $\lambda^{wp}$ and frequency $\nu^{wp}$, whose product is equal to $u^{wp}$.

16. $$\begin{cases} \rho C_V^{wp} = \dfrac{\partial q_T^{wp}}{\partial T} = \dfrac{\partial}{\partial T}(mq_T) = q_T \dfrac{dm}{dT} + m\dfrac{\partial q_T}{\partial T} = q_T \dfrac{dm}{dT} + m\rho C_V = \\ \quad = m\rho C_V \left[ \dfrac{q_T}{m\rho C_V} \dfrac{dm}{dT} + 1 \right] = m\rho C_V \left[ \dfrac{m^*}{m^2} \dfrac{dm}{dT} T + 1 \right] \end{cases}$$

It was demonstrated [22-23] that it holds $0 \le \left[ \dfrac{m^*}{m^2} \dfrac{dm}{dT} T + 1 \right] \le 1$ and therefore $C_V^{wp} \le C_V$; here it is the case to spend few words on the temperature dependence of $m$. At the triple point quite all the energy is supposed to be propagated by means of phonons and wave-packets. As the melting progresses, the solid phase gives progressively way to the liquid dual composition, characterized by the dynamic *icebergs* swimming in the ocean of amorphous liquid. Upon a $T$ increase, the amount of amorphous liquid also increases, while the number and size of *icebergs* decrease, up to get the pure liquid. Therefore, at the triple point one has $m \cong 1$, while $m$ decreases progressively to zero, so that $\dfrac{dm}{dT} < 0$, and the energy propagation by means of wave-packets becomes less and less important as the temperature increases. The evolution of $m$ and how it is modelled in the DML were discussed extensively in [22]. The reasoning is based on the experimental evidence of the presence in liquids of transversal modes [67-86] activated for the propagation of elastic energy by means of shear waves working as in the solid phase. These modes however persist in the liquid phase as long as solid-like structures survive [88,94]. Experiments [67-86] show that the two transversal modes disappear as the system approaches the critical point, where only the longitudinal collective modes survive, accounting for the compression and rarefaction waves responsible for hydrodynamic modes propagation.

Equation (15) represents also the flux of thermal energy carried by a wave-packet propagating over the distance $\langle \Lambda_{wp} \rangle$ and driven by the virtual gradient $\left\langle \dfrac{\delta T}{\delta z} \right\rangle$. Therefore, $j_{+z}^{wp}$ may be re-written using the (mesoscopic) thermal conductivity of the medium $K^{wp}$:

17. $$j_{+z}^{wp} = -K_{+z}^{wp} \left\langle \frac{\delta T}{\delta z} \right\rangle_{+z} = -\frac{1}{6} \left\{ u^{wp} \langle \Lambda_{wp} \rangle \rho C_V^{wp} \right\} \left\langle \frac{\delta T}{\delta z} \right\rangle = -\frac{1}{6} D^{wp} \rho C_V^{wp} \left\langle \frac{\delta T}{\delta z} \right\rangle .$$

Compiling this with eq.(16), one gets:

18. $$K_{+z}^{wp} = \frac{1}{6} D^{wp} \rho C_V^{wp} = \frac{1}{6} u^{wp} \langle \Lambda_{wp} \rangle \frac{\partial}{\partial T} \left[ n^{wp} \langle \varepsilon^{wp} \rangle \right] = \frac{1}{6} u^{wp} \langle \Lambda_{wp} \rangle m\rho C_V \left[ \frac{m^*}{m^2} \frac{dm}{dT} T + 1 \right]$$

By repeating the reasoning for the opposite direction "$-z$", we get an expression like the Eq.(18) with the opposite sign; by summing them, we eventually get:

19. $K^{wp} = \frac{1}{3} u^{wp} \langle \Lambda_{wp} \rangle \frac{\partial}{\partial T} \left[ n^{wp} \langle \varepsilon^{wp} \rangle \right] = \frac{1}{3} u^{wp} \langle \Lambda_{wp} \rangle m \rho C_V \left[ \frac{m^*}{m^2} \frac{dm}{dT} T + 1 \right] = \frac{1}{3} u^{wp} \langle \Lambda_{wp} \rangle \rho C_V^{wp}$

Two considerations about Eq. (19) for the phononic contribution to the thermal conductivity are in order. Firstly, it has exactly the same form as that derived by Boltzmann for the thermal conductivity due to the particle collision and drift in a fluid medium [101, Eq.(86)], the quantity *G* introduced by Boltzmann is here replaced by $\rho C_V^{wp}$; in turn, this depends upon $m$, the number of DoF pertaining to the *particles* involved in the process [101, Eq.(88)]. Second, and consequently to the first, in an isotropic solid, the kinetic theory based on the Boltzmann transport equation (BTE) leads to a thermal conductivity that depends on phonon properties following $K^{BTE} = \frac{1}{3} \int_0^{\omega_m} C(\omega) \upsilon(\omega) \Lambda(\omega) d\omega$, where C is the spectral volumetric specific heat, $\upsilon$ is the group velocity, and $\Lambda$ is the phonon mean free path, all of which depending on the phonon angular frequency ω [102]. Considering average quantities mediated on the spectral range,

20. $K^{BTE} = \frac{1}{3} u^{wp} \langle \Lambda_{wp} \rangle \rho C_V^{wp} = \frac{1}{3} C v_g^2 \tau$ ,

that is the same expression as eq.(19), last member. The reader has certainly recognized that eq.(20) is the general expression of the Debye-Peierls [103] thermal conductivity associated to thermal energy carriers, also known in the form $K^{DP} = \frac{1}{3} n_c u_c \lambda_c c_V$ , where $n_c$ is the carrier number density, $c_V$ the specific heat per carrier, $u_c$ the carrier speed (that effectively participating to the energy propagation) and $\lambda_c$ its mean free path. The Debye-Peierls model accounts for the phonon contribution to thermal conductivity in crystalline solids, in particular explaining how heat is transported by quantized lattice vibrations (phonons) and restricted by anharmonic scattering (Umklapp processes). The general expression is based on the kinetic theory of gases, adapted for phonons, where the thermal conductivity is the summation of contributions from all phonon modes, typically integrated over frequency up to the Debye frequency. These comparisons increase the confidence not only on eq.(19) but generally on all the DML framework.

Let us now analyze more closely the effect of an external temperature gradient $\frac{dT}{dz} = \nabla T^{SS} = \frac{T_1 - T_2}{L}$ applied to the liquid on the current of wave-packets, where SS stays for Steady State and $L$ the extension of the system along $z$. We assume again that $\nabla T^{SS}$ is small enough with respect to $\left\langle \frac{\delta T}{\delta z} \right\rangle$ so that $n^{wp}$ shall not to change significantly, only the average energy of wave packets being affected by the presence of the external disturbance (on this assumption we sill return later in the manuscript). Therefore, non-linear effects can be neglected and the initial state of the system at the microscopic level locally continues to be similar to that at uniform temperature, except for the imbalance of thermal excitations along $z$ [22,24]. Let $\langle \nu_p \rangle$ be the average frequency of "*wave-packet* ↔ *liquid particle*" collisions at $T_0$, with $T_0 \equiv \frac{T_2 + T_1}{2}$ in isothermal conditions due to $\left\langle \frac{\delta T}{\delta z} \right\rangle$. Being it equally present in all the three directions, for each direction becomes $\langle \nu_p \rangle / 6$. In the above hypothesis, if $\nabla T^{SS}$ is applied to the liquid, the total number of collisions per second $\langle \nu_p \rangle / 6$ remains constant but the number of collisions in the direction of heat propagation will increase by $\delta \langle \nu_p \rangle$ per second, and an equal defect, $-\delta \langle \nu_p \rangle$, shall be present in the opposite direction. Therefore a *liquid particle* at a given place in the liquid experiences the same number of collisions per second as if the temperature was uniform, however with an imbalance in the number of collisions with wave-packets between the hotter and the cooler half-spaces along $z$. In the linear range, we have that $2 \cdot \delta \langle \nu_p \rangle$ total excess of collisions, proportional to the temperature gradient $\nabla T^{SS}$. Consequently, for the heat flux $J_q^{ext} = -K_l \nabla T^{SS}$ superimposed to two equal and opposite heat fluxes $+ j_{+z}^{wp} = -K^{wp} \left\langle \frac{\delta T}{\delta z} \right\rangle_+$ and $j_{-z}^{wp} = -K^{wp} \left\langle \frac{\delta T}{\delta z} \right\rangle_-$ it holds:

$$21.\quad \frac{2 \cdot \langle \delta \nu_p \rangle}{2 \cdot \langle \nu_p \rangle / 6} = \frac{\nabla T^{SS}}{\langle \delta T / \delta z \rangle}.$$

The above equation gives the ratio of the frequency of "*wave-packet* ↔ *liquid particle*" collisions due to the applied gradient, to that due to the random motions of collective thermal excitations. It states that macroscopic and microscopic (or mesoscopic) heat conduction obey the same phenomenological relations with identical coefficients. Accordingly, an external temperature gradient $\nabla T^{SS}$ increases by $\frac{\nabla T^{SS}}{\langle \delta T / \delta z \rangle}$ the flux of phonons with respect to that corresponding to the local microscopic heat currents $j^{wp}$ due to spontaneous phonon diffusivity. The clusters will experience $2\delta \langle \nu_p \rangle$ collisions in excess along the direction of the heat flux and will execute as many jumps per second of average length $\langle \Lambda \rangle$ in excess in the same direction. Consequently, every particle travels the distance $2 \langle \Lambda \rangle \delta \langle \nu_p \rangle$ per second along the direction of heat flux generated by $\nabla T^{SS}$ in stationary conditions. This quantity represents the drift velocity $\langle \upsilon_p^{th} \rangle^{SS}$ of the *liquid particle* along $z$ due to $\nabla T^{SS}$:

22. $\left\langle \upsilon_p^{th} \right\rangle^{SS} = 2\langle\Lambda\rangle\langle\delta\nu_p\rangle = \frac{\langle\Lambda\rangle\langle\nu_p\rangle}{3}\frac{\nabla T^{SS}}{\langle\delta T/\delta z\rangle}$.

The above argument shall be adapted and re-used in the next section when dealing with the case of time-dependent, transient phase occurring after the application of an external disturbance; furthermore, it can be used with the obvious adaptations for any other disturbance induced in a system, for instance due to a concentration gradient, as done in a further dedicated section later on.

Before closing this section, we submit to the reader' attention the fact that eqs.(13), (15) and (19) depend, among others, upon the phonon concentration, $n^{wp}$, a quantity that has not been quantified yet. Because the consistency of those equations, as well as of many others equally containing $n^{wp}$ , could be questioned depending upon its value, we procced now to its evaluation. To answer such doubt we may get help from Eq.(4). It provides the variation of the energy density $\delta q_T^{wp} = \delta(mq_T)$ within the liquid along a given direction, equal to the energy shift associated to a Brillouin doublet, $\Delta\omega_B(k) = \pm c_s q$ , multiplied by the phonon density, $n^{wp}\langle\Delta\varepsilon^{wp}\rangle$. Definitively, in water, with a value of 1 calorie for the specific heat and with the same values for the wave-packet frequency as those given in section 2.1, one has

23. $n^{wp}\langle\Delta\varepsilon^{wp}\rangle = n^{wp}h\Delta\langle\nu^{wp}\rangle = \delta(mq_T) = m\rho C_V\Delta T \Rightarrow n^{wp} = \frac{m\rho C_V\Delta T}{\langle\Delta\varepsilon^{wp}\rangle} \approx m\cdot 4\cdot 10^{22}\,cm^{-3}$.

The OoM of the density of phonons, dealt with as *lattice particles*, is therefore of the same OoM as the molecular density (for water the molecular density holds $\approx 4\cdot 10^{22}\,cm^{-3}$ at ambient temperature), or even lesser, because $0 \le m \le 1$. This result supports the hypothesis of neglecting multiple interactions of wave-packets with particles. This numerical result is obtained assuming that the density of phonons does not change by changing the temperature, the only variation being their energy.

### *2.3 Energy Propagation: Cattaneo time-dependent propagation equation "with memory" and the tunnel effect in the DML.*

In this subsection, the application of the DML shall be extended to the non stationary, time-dependent configuration that arises suddenly after the application of a stimulus to a system, affecting its equilibrium. Although the reasoning proposed is valid whatever the nature of the stimulus, it will be specifically applied to the thermal energy propagation, to provide the reader with an immediate comparison with the analogous time-independent approach discussed in the previous subsection. As in the stationary case, equations will be tailored only on the phononic part of the system. Here we propose a short summary of the problem and of the solution offered by the DML approach to provide to the reader with a wide landscape of the capabilities of the DML in representing the possible missing link between the microscopic and macroscopic worlds of the many-body systems. The readers interested to an in-depth discussion of the topic are addressed to [24]. The same hypotheses on the macroscopic system given in the previous subsection are here assumed.

NET equations are typically time-independent, in fact they describe systems in stationary equilibrium. The DML allows to correctly framing also time-dependent phenomena, due to one of its fingerprint, the tunnel effect. Stationary equilibrium in a system following the application of an external stimulus is not achieved instantaneously. The *instantaneous response* typical of the time-independent diffusive phenomena and mathematically described by parabolic equations, should however be replaced by the more appropriate *causality* behavior, accounting for the time-dependent propagative evolution. The DML spontaneously provides hyperbolic equations (of the Cattaneo-Vernotte type, [24]) to describe the time-dependent initial energy propogation [104-111]. However, their mathematical solution falls in contradiction with the rate of entropy production, $\dot{\sigma}$. In fact, although increasing with time in the average, $\dot{\sigma}$ has an undulatory character and consequently, there is an alternation in a system of states out of equilibrium, in which $\dot{\sigma} < 0$. This issue was solved by the formal building of the Extended Irreversible Thermodynamics (EIT) [112], by introducing the new concept of non-equilibrium temperature. How is it faced in the DML framework? The calculation of the entropy within the DML framework shall be the topic of a dedicated paper [113], however we may anticipate that the issue is bypassed in the DML because of its dual character, that is reflected into the tunnel effect and in the possibility of splitting the Lagrangian of a liquid in two equations, one for each of the two components. We will show [113] that the role of the non-equilibrium temperature introduced in the EIT may be represented by the temperature field experienced by the *lattice particles* (or wave-packets) and due to the virtual temperature gradient, $\left\langle \frac{\delta T}{\delta z} \right\rangle$.

How does the tunnelling in the DML work? Its role is of transferring the energy from one point in the liquid to another, exploiting the excitement of the internal DoF of the icebergs, which have own characteristic structural relaxation times (see Table 1 in [22]). Due to such effect, the energy disappears from the heat current for a time lapse coincident with the relaxation time. Therefore, this classical tunnelling allows energy to cover distances otherwise prohibited and to provide a physical interpretation of the delay term in propagative equations. When the wave-packet hits a *liquid particle*, the force $f^{th}$ develops, it acts for $\langle \tau_p \rangle$ seconds displacing the particle by $\langle \Lambda_p \rangle$. Energy $\langle \Delta\varepsilon^{wp} \rangle$ and momentum $\langle \Delta p^{wp} \rangle$, given by Eqs.(5) and (6), respectively, lost by the wave-packet upon the collision, are transferred to the *liquid particle* during $\langle \tau_p \rangle$, increasing the kinetic and potential energy of the last, and exciting its internal vibrational energy levels. The internal DoF oscillate similarly to those pertaining to solid state, giving tise to (quasi) elastic waves with characteristic wavelength $\langle \Lambda_0 \rangle$ and period $\langle \tau_0 \rangle$. $\langle \tau_R \rangle$ seconds later and $\langle \Lambda_R \rangle$ meters forward they are given back to the wave-packet reservoir, and the process is repeated again $\langle \tau_{wp} \rangle$ seconds later and $\langle \Lambda_{wp} \rangle$ meters forward [22,24]. This cascade process is depicted in Figure 5. For each interaction, the length of the tunnel is $\langle \Lambda \rangle$ meters and the energy takes $\langle \tau \rangle$ seconds to cross it. Therefore, $\langle \tau \rangle$ is the time interval during which the energy disappears from the liquid thermal reservoir for being stored in the internal DoF of the *liquid particles*; once $\langle \tau \rangle$ has elapsed, the energy reappears $\langle \Lambda \rangle$ meters forward. The DML provides a physical need for the relaxation times, introduced *ad hoc* by Maxwell and Frenkel, and allows for solving from the physical point of view the dilemma of infinite diffusion velocity typical of diffusive-like equations.

The Cattaneo-like equation for heat propagation given in the DML framework [24],

24. $$J_q^{wp} + \langle \vartheta \rangle \frac{\partial J_q^{wp}}{\partial t} = -K^{wp} \frac{\partial T}{\partial z},$$

is characterized by the presence of the time derivative of the heat flux, $\frac{\partial J_q^{wp}}{\partial t}$. Combining eq.(24) with the usual continuity equation, one easily gets the hyperbolic propagation equation in terms of temperature $T$:

25. $$\frac{\partial^2 T}{\partial z^2} = \frac{\rho C_p^{wp}}{K^{wp}} \frac{\partial T}{\partial t} + \langle \vartheta \rangle \frac{\rho C_p^{wp}}{K^{wp}} \frac{\partial^2 T}{\partial t^2}$$

Equations (24) and (25) hold for the quantities related to the collective excitations, with $\langle \vartheta \rangle = n \langle \tau \rangle$, $n \geq 1$ [24], and $K^{wp}$ given by eq.(19) ($\langle \tau \rangle$ has been replaced by $\langle \vartheta \rangle$ due to the randomization process). The additional term $\langle \vartheta \rangle \frac{\partial J_q^{wp}}{\partial t}$ in eq.(24) is the total energy subtracted from the collective excitations reservoir; part of it becomes kinetic energy of *liquid particles*. Dividing it by $u^{wp}$, one gets the momentum flux $J_p^{wp} = \frac{1}{u^{wp}} \langle \vartheta \rangle \frac{\partial J_q^{wp}}{\partial t}$ transferred from the *lattice particles* to the *liquid particles*. Dimensionally, $J_p^{wp}$ is also a pressure, $\Delta P^{wp} \equiv J_p^{wp}$, to which the liquid responds with the self diffusion of the *liquid particles*. An expression analogous to eq.(25) may be easily obtained for the heat flow $J_q^{wp}$, for a mass concentration or even for a matter flow, in case one would consider the mass propagation instead of the heat. What really matters here is that the intrinsically negative term $\frac{\partial J_q^{wp}}{\partial t}$, added by Cattaneo and Vernotte on pragmatic or logical reasoning, finds a physical interpretation in the DML framework. It is exactly this term which accounts for that part of the energy temporary stored into the *liquid particles* for the time interval $\langle \tau \rangle$, i.e. the *relaxation time,* in a non propagating form (Figure 4). During such time-lapse, the linear relationship between the heat flux and the temperature gradient does no longer hold, the presence of $\frac{\partial J_q^{wp}}{\partial t}$ making $J_q^{wp}$ and $\nabla T$ no longer proportional to each-other everywhere and every time. However, it is important to highlight the following point: both the interactions and the relaxation mechanisms are always operating in a liquid whatever its thermodynamic state; therefore they are present not only during the transient phase, but also when the system is at equilibrium, or with an external temperature gradient at the steady state. The differences between these two cases are that i) at thermal equilibrium, these interactions are equally likely to occur along any direction, so that they do not have any effect at macroscopic level; ii) when the stationary equilibrium is attained, again they have no longer effect on the heat propagation at macroscopic level because they are ubiquitous in the system, as the thermal current has established and stabilized along the entire system. Said in other words, every transient effect has been overcome in the dynamics of the system at the macroscopic level. The Cattaneo propagation equation (24) reduces to the classical Fourier law, eq.(11), when the time elapsed from the beginning of the heat propagation has become large enough to make negligible the contribution of the additional term, being it due to the tunneling, that has no longer effect on macroscopic scale.

We want now dedicate some words to the interpretation in the DML framework of several consequences of Equations (24) and (25). Equation (24) is the simplest equation combining diffusion and waves, giving rise to propagation and, not secondary, allowing for dissipation to take place. It states that there is a physical mechanism, the tunnel, by means of which energy is not freely flowing through the liquid, but is temporarily subtracted for a time $\langle \vartheta \rangle$. This affects only the transient phase, (although still present also at the steady state), because it becomes irrelevant when a time interval larger than $\langle \vartheta \rangle$ has elapsed. Hence $\langle \vartheta \rangle$ represents the finite build-up time for the onset of a thermal current once a temperature gradient is applied to a system. The heat does not start to flow instantaneously but rather grows gradually with the delay time $\langle \vartheta \rangle$. Conversely, if a thermal gradient is suddenly removed from a system at stationary equilibrium, there is a lag in the disappearance of the heat current, and eq.(24) exhibits just such a delay [114].

Equations like (24) or (25) are encountered in physics in the several areas where two subsystems belonging to the same closed system mutually interact. They yield complex solutions (frequencies), but only those with non-zero real parts are representative of propagating modes, of interest here. To the best of the author' knowledge, this is the first attempt to use the telegrapher equation to describe the propagation of phonons in normal liquids as carriers of the thermal (and elastic) signal[2]. Indeed it is precisely the presence of the two interacting populations, *lattice particles* and *liquid particles*, to justify the presence of the relaxation term in the equation for heat transport (24), that we have proposed to be linear in the relaxation time. The tunnel effect characterising these interactions is the physical mechanism allowing the interpretation of the heat transport to be of a dynamical nature, thus allowing escaping from the problem of the infinite diffusion velocity. Incidentally, the fact that the tunnel effect produces a mesoscopic lattice redistribution of the energy content is reminiscent of the Interstitial Theory of Granato [116] for the liquid specific heat. In his model, Granato determined that the linear temperature-dependent component of the specific heat at low temperature is identified as occurring due to the interstitialcy tunneling between different orientations of the interstitialcies themselves.

The classical treatment of heat propagation involves only the molecular part through the mutual interactions of molecules; therefore, the hyperbolic-type equations were believed not easily adaptable to describing the heat propagation because of the lacking of physical mechanism accounting for the temporary subtraction of the energy from the heat current during the transient phase. In the DML framework this limitation is byassed because that part of energy propagating by means of the interactions of wave packets with the pseudo-crystalline structures, is described by a Cattaneo-like equation with "memory". If, on the one hand, classically it is not possible to provide an explanation either of the delay term or of the undulatory behaviour in the transient phase, on the other hand, in the DML, the delay term reveals the physical relevance of the tunnelling, and the undulatory behavior is a consequence of the presence of the phonons and of their role as energy carriers. However, a formal generalization of the propagation equation, including either the part related to the wave-packets (*wp*) or the "classical" one related to molecular interactions (*mol*), now mutually interacting, may be advanced, namely:

$$26. \quad \begin{cases} mol: \; J_q^{mol} + \left\langle \vartheta^{mol} \right\rangle \dfrac{\partial J_q^{mol}}{\partial t} = -K^{mol} \dfrac{\partial T}{\partial z} \\ wp: \; J_q^{wp} + \left\langle \vartheta^{wp} \right\rangle \dfrac{\partial J_q^{wp}}{\partial t} = -K^{wp} \dfrac{\partial T}{\partial z} \end{cases}$$

each one with own relaxation time and thermal conductivity. This scheme copes with the assertion of many authors [61-63,117] that liquids exhibit a distribution of relaxation times rather than a single one. Because the internal energy of the system is distributed between the two sub-systems, a single energy balance equation for the whole system should be considered. Equation (26) *mol* is associated with the macroscopic sound velocity, while Equation (26) *wp* is associated with the hyperfrequency sound velocity detected at mesoscopic scale [73]. However, what really matters is that the thermal evolution of the system is characterized by two wave equations, each one with its own medium parameters, relaxation time and wave propagation velocity. Indeed, because of the dualism, the situation in the DML recalls the one proposed by Landau for the superfluidity of HeII [55], i.e., that below the $\lambda$ –point liquid He is composed both of ordinary and superfluid He, each one with its own sound velocity.

[2] Interestingly, in 1946 Peshkov [115] had hypothesized that in low temperature liquids "*a gas of thermal quanta capable of performing vibrations similar to those of sound should exist*".

The separation of the two contributions participating to the liquid dynamics has a strict correlation also with the emergence of a gap in the ***k***-space, that is in turn correlated with hyperbolic equations and with systems made of two interacting subsystems. The "***k***-*gap*" or *Gapped Momentum States* (GMS) [118], is crucial in the identification of the relaxation times typical of a dual system. In their intriguing approach, Baggioli et al. [53,119] build up the Lagrangian $\phi_0$ describing systems exhibiting ***k***-gap and constituted by two mutually interacting sub-systems. Following [53,118-120], let $\phi_1$ and $\phi_2$ be the two fields of $\phi_0$ representing displacements and velocities of the particles of a medium. The authors show that the equations of motion for the two scalar fields decouple, leading to two separate Cattaneo-like equations for $\phi_1$ and $\phi_2$, whose expressions are

$$27.\quad \begin{cases} \phi_1 = \phi_0 \exp\left(-\dfrac{t}{2\langle\tau\rangle}\right)\cos(kx-\omega t) \\ \phi_2 = \phi_0 \exp\left(\dfrac{t}{2\langle\tau\rangle}\right)\cos(kx-\omega t) \end{cases}$$

The interaction potential $\phi_0$ is an oscillating function with period $\langle\tau\rangle$ (see Figure 5 in [118] and Figure 2 in [53]), i.e. $\phi_1$ and $\phi_2$ alternatively reduce and grow over $\langle\tau\rangle$. The physical interpretation of this behaviour is that the two interacting sub-systems, represented by the two scalar fields, exchange among them energy and momentum, like wave-packets and *liquid particles* do in the DML. Because it holds $\phi_0 = \phi_1 \cdot \phi_2$, the total energy of the whole system does not vary with time, i.e. it is a constant of motion, as expected in the DML for systems in stationary equilibrium, constituted by the two populations of mutually interacting sub-systems. Let's therefore return to the eq.(25), whose solutions with non-zero real parts corresponding to propagative modes are

$$28.\quad T(z,t) = T_0 e^{-(1/2\langle\vartheta\rangle)t} e^{i(kz\pm\omega_C t)}.$$

The associated algebraic equation is $\omega^2 + \frac{i}{\langle\theta\rangle}\omega - \upsilon_C^{wp\,2} k^2 = 0$ with $\upsilon_C^{wp} = \sqrt{\frac{K^{wp}}{\rho C_p^{wp}\langle\vartheta\rangle}} = \sqrt{\frac{D^{wp}}{\langle\vartheta\rangle}}$,

where $D^{wp} = \frac{K^{wp}}{\rho C_p^{wp}\langle\vartheta\rangle}$ is the liquid thermal diffusivity. Incidentally, it represents also the phonon dispersion relation. By using the expression for $K^{wp}$ from eq.(19), one gets

$$29.\ \upsilon_C^{wp} = \sqrt{\frac{u^{wp}\langle\Lambda_{wp}\rangle}{3\gamma^{wp}\langle\vartheta\rangle}}$$

with $\gamma^{wp} = C_p^{wp} / C_V^{wp}$. This expression of $\upsilon_C^{wp}$ allows to recognize that it depends on both the relaxation time and the flight parameters of the wave-packets, as one would have expected. The condition to have propagative waves as solutons is:

$$30.\ k > \frac{1}{2}\sqrt{\frac{\rho C_p^{wp}}{K^{wp}\langle\vartheta\rangle}} = \frac{1}{2}\sqrt{\frac{1}{D^{wp}\langle\vartheta\rangle}} = k_m$$

or alternatively:

$$31.\ \langle\vartheta\rangle < \frac{\rho C_p^{wp}}{4K^{wp}k_m^2} = \frac{1}{4D^{wp}k_m^2} = \langle\vartheta\rangle_M .$$

Equation (30), or (31), show us that the dynamics described works above a minimum value for the wave vector, $k_m$, or a maximum for the relaxation time, $\langle\vartheta\rangle_M$. Said in other words, heat propagation is inhibited for waves with wave vector below $k_m$. Therefore, there is a ***k***-gap for a dual system of interactiong particles, such those described by the DML. The ***k***-gap is always present in all the systems where the propagation of energy is described by equations such as eq.(24) or eq.(25) [53,118-120]. Equation (31) represents also the operative definition of $\langle\vartheta\rangle_M$, the maximum relaxation time, once $k_m$ is known from experiments.

Another peculiarity of eqs.(24) or (25) is that they take into account the capability of a medium to dissipate the thermal energy carried by the wave packets. This circumstance is reflected by the finite propagation range of the thermal wave, in turn due to the presence of $\langle\vartheta\rangle$ with a finite value and different from zero. Equation (31) may be commuted in the following expression

$$32.\ \langle\Lambda_C\rangle = \upsilon_C^{wp}\langle\vartheta\rangle < \sqrt{\frac{K_l^{wp}\langle\vartheta\rangle}{\rho C_p^{wp}}} = \sqrt{D^{wp}\langle\vartheta\rangle} \Rightarrow \langle\Lambda_C\rangle_M = \sqrt{D^{wp}\langle\vartheta\rangle_M} ,$$

where the gap is given in terms of wavelength instead of momentum. The last equality recalls the well-known Einstein relation [121]. Using again $K^{wp}$ from Equation (19) into Equation (32), we get an interesting expression of $\left\langle \Lambda_C \right\rangle_M$ in terms of the relaxation time and of of wave-packet parameters:

33. $$\left\langle \Lambda_C \right\rangle_M = \left\langle \Lambda_{wp} \right\rangle \sqrt{\frac{\left\langle \vartheta \right\rangle_M}{3\gamma^{wp} \left\langle \tau_{wp} \right\rangle}}$$

In [24,119] it was concerned about the amplitude of ***k***-gap; rather, we shift here the attention to the $\langle \vartheta \rangle$-gap or the $\langle \Lambda_C \rangle$-gap, defined in eqs.(31) and (32), respectively. Alternative readings of these equations tell us that there is an upper limit also for the maximum length of the tunnel, or for the maximum number of *liquid particles* with which a *lattice particle* may interact before being definitively damped (remember that it holds that $\langle \vartheta \rangle = n \langle \tau \rangle$ with $n \geq 1$, and that a wave is well defined only if its wavelength is smaller than its propagation distance). This offers us the possibility to illustrate the dynamics occurring in a liquid during the thermal transient. When an external temperature gradient is applied, the thermal content obtained by Equation (4) increases, determining an imbalance of the phonon flux, so that there will be an excess of interactions as well of the energy transferred from the thermal current to the liquid particles through events such as those of Figure 4a. Let us start at $z = z_0$. Each interaction lasts $\langle \tau \rangle = \langle \tau_p \rangle + \langle \tau_R \rangle$, the time interval during which the energy disappears as liquid free energy to become an iceberg's internal and kinetic energy, and the liquid particle moves by $\langle \Lambda \rangle = \langle \Lambda_p \rangle + \langle \Lambda_R \rangle$ forward; after that, the emerging wave packet has lost part of its initial polarization. Depending on the residual energy and momentum, it may interact with another liquid particle $\langle \tau_{wp} \rangle$ seconds later, and the above process is replicated, say, $n$ times. The overall duration of the randomization process lasts $\langle \vartheta \rangle = n \langle \tau \rangle$, during which the particle and the thermometric front will have advanced by $\langle \Lambda_C \rangle = \upsilon_C^{wp} \cdot n \cdot \langle \tau \rangle$, determining a temperature increase by $\Delta T$ over $\langle \Lambda_C \rangle$, so that $\left\langle \frac{\delta T}{\delta z} \right\rangle \approx -\frac{\Delta T}{\langle \Lambda_C \rangle}$. At the end of $\langle \vartheta \rangle$, a liquid warmer both in its molecular and gas of excitation components will be in contact with the still unperturbed medium laying beyond $z = z_0 + \langle \Lambda_C \rangle$, setting the stage for a replica of the events. At $z = z_0$, instead, in absence of a new advancing front, the process of heat propagation after $\langle \vartheta \rangle$ seconds reaches the steady state. Depending on the value of $n$, and compiling $\langle \vartheta \rangle_M$ and $\langle \Lambda_C \rangle_M$ with $\langle \tau \rangle$ and $\langle \Lambda \rangle$ from Equations (5) and (6), respectively, (or with their multiples, $\langle \vartheta \rangle = n \langle \tau \rangle$ and $\langle \Lambda_C \rangle = n \langle \Lambda \rangle$, with $n \geq 1$), one may evidence the following intervals for the propagating modes: (i) the momentum carried by wave packets is too low to interact with *liquid particles*, and there are no propagating modes ($n = 0$); (ii) the momentum is large enough to allow the wave packet to interact with several *liquid particles*, thus obtaining damped oscillating modes and $n \geq 1$; (iii) this is what happens inside a single liquid particle, a solid-like structure with the dispersionless propagation of elastic perturbations. In the last case, one experiences the propagation velocity of elastic waves similar to that of the corresponding solid [22,67-88]. These limitations are therefore an indirect evidence of the presence of pseudo-crystalline structures in liquids, in particular [88,91-93], where the authors recognize the presence of solid-like structures in liquids for distances where purely elastic non-dissipative modes are allowed, a picture absolutely similar to that of the DML.

A more phenomenological interpretation of eqs.(27) goes through the explicitation of the physical meaning of $\phi_1$ and $\phi_2$. The energy lost by a wave-packet is commuted into kinetic (i.e. translational) and potential (i.e. collective vibrations) energies of the iceberg, each process being characterized by a relaxation time. The kinetic part of the particle' energy, involving external DoF, will be dissipated against the liquid friction with a own decay time $\tau_k$, along the decay process

34. $\Delta E_k^p(t) = \Delta E_k^p(t=0) \cdot e^{-t/\tau_k}$.

The part responsible for the excitation of the internal collective vibratory quantized DoF will be relaxed along with the relaxation time(s) $\tau_{\Psi_i}$ pertaining to the several excited distinct DoF, and following the decay processes [117]

35. $\Delta \Psi_i^p(t) = \Delta \Psi_i^p(t=0) \cdot e^{-t/\tau_{\Psi_i}}$.

This event is repeated many times as in a chain (Figure 5), giving rise to the thermal avalanche. To the purpose of the present paper, however, we still consider here a single relaxation time $\langle \tau \rangle$ (or $\langle \vartheta \rangle$), working for our interactions because what is relevant here is not just to distinguish between the several relaxation times one may consider, but rather establish that *there is* a delay physically associated to the *lattice particle* $\leftrightarrow$ *liquid particle* interaction at the origin of $\langle \vartheta \rangle$. This fact *per se* represents one of the novelties of this model, because the delay term in the transient phase in the DML represents a physical consequence of the model.

Let's close this section by evaluating how the average number of *wave-packet* $\leftrightarrow$ *liquid particle* collisions in the Transient Phase (TP) is affected by the external gradient, as done for the SS case. Therefore, we will make use of the same reasoning used for the steady state. Let $\nabla T^{TP}(z,t)$ be the temperature gradient that is esatblishing across the system, $T_h$ the temperature of the heat source applied to the system and $T_c(z,t)$ the temperature of the advancing thermal front at the point $z$ and at time $t$. Because the thermal front advances along the system, it will be:

36. $\nabla T^{TP}(z) = \dfrac{T_h - T_c(z)}{\Delta z(t)}$.

Due to the application of the external temperature gradient, the heat flux crossing the system will give rise to an increase of the number of *wave-packet* $\leftrightarrow$ *liquid particle* collisions in the direction of $J_q$. In fact, in this case, events of type a) in Figure 1 have a larger probability to occur than events of type b). At equilibrium the flux of wave-packets is driven by $\left\langle \frac{\delta T}{\delta z} \right\rangle$ [22]. If $\left\langle \nu_p \right\rangle$ is the average number per second of *wave-packet* $\leftrightarrow$ *liquid particle* collisions due to $\left\langle \frac{\delta T}{\delta z} \right\rangle$, along the direction of $J_q$ such number is $\left\langle \nu_p \right\rangle / 6$ for symmetry reasons. When $\nabla T^{TP}$ is applied, there will be an increase of the frequency of collisions along $z$, let it be $\delta \left\langle \nu_p \right\rangle$ . $\delta \left\langle \nu_p \right\rangle$ is a quantity neither constant in time nor in space, because such is $\nabla T^{TP}$ (see eq.(36)). To evaluate $\delta \left\langle \nu_p \right\rangle$ we make the same hypothesis of the SS discussed before, i.e. that as first approximation the application of $\nabla T^{TP}$ increases proportionally the frequency of *wave-packet* $\leftrightarrow$ *liquid particle* collisions:

$$37. \quad \frac{\left\langle \delta \nu_p \right\rangle}{\left\langle \nu_p \right\rangle / 6} = \frac{\nabla T^{TP}(z)}{\delta T / \delta z}$$

Consequently, every *liquid particle* will execute as many jumps per second in excess as $\delta \left\langle \nu_p \right\rangle$ along the direction of $J_q$, each of average length $\left\langle \Lambda \right\rangle$, for a total distance travelled $\left\langle \Lambda \right\rangle \cdot \delta \left\langle \nu_p \right\rangle$ per second along the direction of heat flux generated by $\nabla T^{TP}$. This quantity represents the drift velocity $\left\langle \upsilon_p^{th} \right\rangle^{TP}$ of the *liquid particle* during the *TP* along $z$ induced by the external temperature gradient:

$$38. \quad \left\langle \upsilon_p^{th} \right\rangle^{TP} = \left\langle \Lambda \right\rangle \left\langle \delta \nu_p \right\rangle = \frac{\left\langle \Lambda \right\rangle \left\langle \nu_p \right\rangle}{6} \frac{\nabla T^{TP}(z)}{\left\langle \delta T / \delta z \right\rangle}.$$

The reader must not be fooled by the comparison of Eq. (38) with the analogous expression (22) for the *SS*, from which one could hastily and erroneously conclude that in the *SS* speed and imbalance of impacts are greater than in the *TP*. In fact, the value of $\left\langle \upsilon_p^{th} \right\rangle^{TP}$ in (38) is a function of $\nabla T^{TP}$, which in the *TP*, as is known, can have much greater values than that of the stationary state, being it applied over much shorter distances. At steady state $\left\langle \upsilon_p^{th} \right\rangle^{TP}$ will assume the value $\left\langle \upsilon_p^{th} \right\rangle^{SS}$ given by eq.(22). It is obious that the previous reasoning also holds true if the flow of heat within the system is due to the presence of a heat sink at a temperature $T_c$ lower than that of the system, or even in the case in which the system is in contact with two different sources, one with a temperature higher than that of the system and the other with a lower (provided that the average temperature of the system is unchanged).

## 2.4 Mass Transport: the case of the thermodiffusion and the Soret dynamical equilibrium in the DML

The previous sections were dedicated to the description of the energy transport in liquids in the framework of the DML. This section is dedicated at discussing a typical coupled transport in liquids, i.e. when an external perturbation affects energy and mass transport. Such phenomena are well known in NET [3-8]; examples are the thermodiffusion, also dubbed thermoforesis or Soret effect, in which the transport of mass is associated to that of thermal energy when a temperature gradient applied to a liquid mixture; the Dufour effect, similar to thermodiffusion, although induced by a concentration gradient; the Peltier effect, manifesting when a voltage difference across a metallic junction gives rise to a thermal flux associated to an electric current; the Seebeck effect, similar to the Peltier, but due to a temperature difference working across a metallic junction, generating an electric current across the junction, etc. Because the DML has a strong impact on the dynamics of molecules and *lattice particles*, we will focus our attention on those effects where only forces of mechanical origin are at stake. The thermodiffusion in a liquid solution is an example. Let us then consider Figure 6, where the two chemical species, solute and solvent, are represented by spheres and cubes[3]. The phenomenon is well known from experiments [see 122-149 for a very limited literature wrt that available worldwide] to the point that it was studied also in microgravity environment [see for instance 150-165]. The setup consists in a temperature gradient imposed to a confined homogeneous binary liquid misture. Provided that several accurate experimental preconditions are applied, one can observe that one of the two species will move towards the cold side, and, according to the Newton' third law, the other is pushed back towards the hot side, thus balancing the concentration gradient that is being generated. This dynamics, of course, does not produce the complete separation of the two chemical species but it reaches a stationary equilibrium. In fact, as soon as the concentration gradient begins to form, the ordinary diffusion begins too, for both the chemical species, opposing to the diffusion driven by the thermal gradient. The thermodiffusion stops at the steady state, named the Soret equilibrium [5-8,122-123,166], representing the dynamical equilibrium between the diffusion due to the thermal waves and working to separate the two chemical species, and the ordinary diffusion, working to restore the homogenous concentration in the liquid mixture. At the Soret equilibrium, one may observe a temperature gradient (external constraint) responsible for a heat flow, and two opposite small concentration gradients (a second order crossed effect), for the spheres and the cubes. Such dynamical equilibrium is described by means of the Soret coefficient $S_T$, that is (conventionally)

[3] Although the Soret effect is known to occur also in multicomponent systems, we will focus our attention on two-components systems for simplicity

positive when the solute, i.e. the less concentrated chemical specie, drifts towards the cold side, while it is negative when it drifts towards the hot one. When the initial condition is represented by a gradient of concentration in a fluid mixture, i.e. a gradient of chemical potential, the system spontaneously evolves towards the equilibrium due to the ordinary Fick diffusion. A careful evaluation of the thermal evolution of the system shows that the spontaneous diffusion of the chemical species of a mixture generates a temperature gradient: this is the Dufour effect in NET. Despite the weakness of such effects, their study in thermodynamics is motivated by the fact that coupled heat and mass fluxes occur in many relevant processes, such as distillation, extraction, crystallization, absorption, drying or even condensation, the thermodiffusion can play an important role in many natural and industrial processes [167-171]. We will shortly recall in this section the interpretation of the Soret equilibrium and the thermodiffusion within the framework of the DML, the more extensive and complete treatment being given in [18]. The reader certainly understands that the reasoning may be easily extended to other crossed phenomena typical of NET.

To come into the details of the physical model, we start from the set of equations (1). Considering temperature and concentration gradients as generalized forces and heat and mass current as generalized fluxes, tailored for a 2-components fluid system, it becomes [3-8]:

39. $$J_q = -L_{qq}\frac{\nabla T}{T^2} - L_{q1}\left(\frac{\partial \mu_1}{\partial w_1}\right)_{T,p}\frac{\nabla w_1}{(1-w_1)T}$$

40. $$J_m^1 = -L_{1q}\frac{\nabla T}{T^2} - L_{11}\left(\frac{\partial \mu_1}{\partial w_1}\right)_{T,p}\frac{\nabla w_1}{(1-w_1)T}.$$

The physical quantities now introduced are the mass flux $J_m^1$, the chemical potential and mass fraction of component 1 (the solute for instance), $\mu_1$ and $w_1$, the phenomenological Onsager coefficients describing the proportionality between the generalized forces and the fluxes they generate, $L_{1q}$, $L_{q1}$, $L_{11}$. In particular, $L_{1q}$ and $L_{q1}$ are the cross-correlated phenomenological coefficients, for which $L_{1q} = L_{q1}$ holds in the Onsager framework (reciprocity postulate). Equations (39) and (40) are often re-written in terms of experimental coefficients instead of the Onsager' ones, giving the following constitutive equations:

41. $$J_q = -K\nabla T - \rho w_1\left(\frac{\partial \mu_1}{\partial w_1}\right)_{T,p} T D_T^D \nabla w_1$$

42. $$J_m^1 = -\rho w_1 w_2 D_T^S \nabla T - \rho D \nabla w_1.$$

Again, the new physical quantities introduced are the ordinary Fick diffusion coefficient, $D$, the Soret $D_T^S$ and the Dufour $D_T^D$ thermal diffusion coefficients, respectively; $K$ and $\rho$ refer here to the liquid mixture. The Soret equilibrium is attained when the thermal profile is stabilized through the medium, the heat still flows through it and the thermodiffusion is balanced by the ordinary mass diffusion, so that $J_m^1 = 0$ in Eq.(42). This last condition gives the definition of $S_T$:

$$43.\ S_T = \frac{D_T^S}{D} = -\frac{1}{w_1 w_2}\left(\frac{\nabla w_1}{\nabla T}\right)_{J_m^1=0}$$

The previous argument also clarifies, among others, that the Soret equilibrium is not a static equilibrium, but a stationary equilibrium, because it is reached when two opposite mass currents balance perfectly each other. As such, we will observe, even when the equilibrium is maintained, a positive rate of entropy production [166].

In order to get $S_T$ in the framework of DML, we now take in consideration the duality of the liquid. Let's start of course from the physical interpretation of the thermodiffusion, for which we make reference to Figure 1. The thermodiffusion begins with a homogeneous mixture crossed by a heat flux due to an external thermal gradient. Therefore, there is a wave-packet current in excess from the hot to the cold side; this pushes the molecules, for instance the "1" of Figure 6 – we will see precisely "which" – following the mechanism of Figure 1a. Consequently, as far as the phenomenon goes on, they accumulate in the cold zone, giving rise to a concentration gradient. The further consequence is that, as far as the concentration gradient increases, the ordinary diffusion of molecules "1" takes place, giving rise to an opposite current. The phenomenon reaches the Soret stationary equilibrium when the two mass currents are equal each other on the average. Another consequence, unnoticed in textbooks but that is clearly evident in the DML framework, is that the diffusive countercurrent gives rise also to a faint thermal gradient opposing to the external one, due to a Dufour effect occurring while the concentration gradient is establishing, and being due to the occurrence of processes following Figure 1b.

Having clarified such point, let now $D_p^{th}$ and $D_p$ be the *liquid particle* diffusion coefficient in a thermal gradient and that of ordinary diffusion, respectively. For both we will calculate the expression due to the collision with the *lattice particles*. Equation (22) allows calculating $D_p^{th}$, i.e. the particle drift in a unitary temperature gradient, that we assume in DML be the equivalent of $D_T^S$ defined in NET:

$$44.\ D_p^{th} = \frac{\langle \upsilon_p^{th} \rangle}{dT/dz} = \frac{\langle \Lambda \rangle \langle \nu_p \rangle /3}{\langle \delta T/\delta z \rangle}$$

This expression should not surprise us; in fact, $D_p^{th}$ is a signature of the "solute+solvent" mixture, therefore it is correctly given in terms of the *liquid particle* drift $\langle \Lambda \rangle$ caused by the collisions $\langle \nu_p \rangle$ with the *lattice particles*. Even more relevant is its (reciprocal) dependence upon $\left\langle \frac{\delta T}{\delta z} \right\rangle$, being it also responsible in eq.(44) for the algebraic sign, determining in such a way the possibility for $D_p^{th}$ to be either positive or negative, as expected from theory or experiments. The Soret coefficient $S_T$ is now obtained from its definition given by eq.(43), exploiting the Einstein relation for $D_p$ (we supposed the mixture sufficiently dilute so that the molecules, either of solute or of solvent, do not interfere each other during their diffusion):

$$45.\ S_T = \frac{D_p^{th}}{D_p} = \frac{\langle\Lambda\rangle\langle\nu_p\rangle}{3\langle\delta T/\delta z\rangle} \Bigg/ \frac{\langle\Lambda\rangle^2\langle\nu_p\rangle}{3} = \frac{1}{\langle\Lambda\rangle\cdot\langle\delta T/\delta z\rangle}.$$

The algebraic sign of $S_T$ is hidden in the term $\left\langle \frac{\delta T}{\delta z} \right\rangle$. We have extensively discussed the consequences and implications of eq.(44) and eq.(45) in [18], with particular attention to the sign inversion.

A final remark is in order on eq.(45): recalling that $\langle\Lambda\rangle$ is the *liquid particle* drift caused by the $\langle\nu_p\rangle$ collisions per second with the *lattice particle*, it becomes:

$$46.\ S_T = \frac{1}{\langle\Lambda\rangle\cdot\langle\delta T/\delta z\rangle} \approx \frac{1}{\langle\delta T\rangle_z}.$$

Here $\langle\delta T\rangle_z$ is the average temperature difference experienced by a solute *liquid particle* over the distance $\langle\Lambda\rangle$ along $z$ covered after a collision with a phonon. This is a direct consequence of the dynamics characterizing the DML and fully in line with the macroscopic meaning of the Soret coefficient in NET. It is indeed not surprising getting a sign dependence for $S_T$, or for $D_p^{th}$, in the DML framework. The origin of such a characteristic resides in the theoretical formulation of the thermal force $f^{th}$ (see Figure 1),

$$47.\ f^{th} = \sigma_p \delta\left(\frac{J_q^{wp}}{u_\varphi}\right) = \sigma_p \delta\Pi^{wp}$$

that one may deduce from first principles, as shown in [18,22]. An alternative expression for $S_T$ can be obtained starting from that of the pressure gradient exherted by the current of wave-packets travelling through the liquid medium, $\nabla\Pi^{wp}$, in turn equivalent to the gradient of the energy density $q_T^{wp}$ given by eq.(4). In eq.(47), $u_\phi$ is the phase velocity associated with the wave-packet and $\sigma_p$ is the cross-section of the "obstacle", the solid-like cluster for instance, on the surface of which $f^{th}$ is applied. Equation (47) tells us that it is now $f^{th}$ that can either be positive or negative, depending on whether $\frac{J_q^{wp}}{u_\varphi}$ increases or decreases upon the interaction with the obstacle. Very instructive and helpful is the concept of Radiant Vector, for the definition of which the reader is again addressed to [18]. Exploiting such a concept and applying it to the momentum transferred by thermal waves impinging on a solute molecule in a liquid mixture, we get:

$$48.\ \nabla\Pi^{wp} \equiv \nabla\Pi^{th} = \nabla\left(\frac{J_q^{wp}}{u_\varphi}\right) = -\psi^{th}$$

$\psi^{th}$, dimensionally a force per unit of volume, is responsible for momentum transport associated with the wave-packets propagation due to the pressure gradient. Consequently, it leads to the appearance of the force $f^{th}$ acting on molecular clusters. By rewriting Eq.(48) in terms of discrete quantities, one yields the following expression for the net $\Delta\Pi^{th}$:

$$49.\ \Delta\Pi^{th} = \left[\left(\frac{J_q^{wp}}{u_\varphi}\right)_1 - \left(\frac{J_q^{wp}}{u_\varphi}\right)_2\right] = \frac{f^{th}}{\sigma_p} \quad \Rightarrow \quad f^{th} = \sigma_p\left[\left(\frac{J_q^{wp}}{u_\varphi}\right)_1 - \left(\frac{J_q^{wp}}{u_\varphi}\right)_2\right]$$

The time averaged pressure generated by the ubiquitous wave-packets stream in a system without external gradients shall be obviously zero. This is no longer true if an external thermodynamic force works on the system. Assuming that the external gradient is small enough with respect to the virtual gradient, so that linear deviations suffice to describe the induced variations due to its application, eq.(21) still holds for $\langle\delta\nu_p\rangle$. This increase of the number of collisions causes, in turn, an increase of the energy and momentum exchanged upon the collisions, namely:

$$50.\ \delta W^{th} = \langle\delta\nu_p\rangle \cdot \Delta\varepsilon^{wp}$$

$$51.\ \delta F^{th} = \langle\delta\nu_p\rangle \cdot \Delta p^{wp}\ .$$

$\delta W^{th}$ and $\delta F^{th}$ are the power dissipated and the force exerted on the collision target, respectively, due to the surplus $\langle\delta\nu_p\rangle$ of collisions. At the Soret equilibrium, $\delta W^{th}$ is the power needed to sustain the thermal and concentration gradients of the mixture (the Soret equilibrium is characterized by a positive rate of entropy production, the ratio $\delta q/T$ at the cold side being higher than that at the hot side, $\Delta s = \delta q(1/T_c - 1/T_h) > 0$), while $\delta F^{th}$ is the net force responsible for the separation of the two chemical species, for instance cubes and spheres of Figure 6.

Compiling eqs.(21), (47) and (49), and adapting eq. (51) to the "discrete" situation involving the two distinct *liquid particles* "1" and "2", we get that:

$$52.\ \delta F^{th} = \sigma_p \cdot \Delta\Pi^{th} = \frac{\langle\nu_p\rangle}{6}\frac{dT/dz}{\langle\delta T/\delta z\rangle}\Delta p^{wp} = \sigma_p \cdot \Delta\left(\frac{J_q^{wp}}{u_\phi}\right) = \sigma_p\left[\left(\frac{J_q^{wp}}{u_\phi}\right)_1 - \left(\frac{J_q^{wp}}{u_\phi}\right)_2\right]$$

This equation tells us "what" of the two species, "1" or "2", is pushed by the $\langle\delta\nu_p\rangle$ collisions to the cold side and "why": the chemical species which is pushed towards the cold side by the external thermal gradient is the one for which the difference $\Delta\left(J_q^{wp}/u_\phi\right)$ is positive, and therefore the sum of the two vectors is a greater than every single vector. Of course, the opposite happens for the other chemical species. Because the quantity usually observed in solutions is the concentration of the solute, i.e. the less concentrated, one refers to positive or negative Soret coefficient depending on its concentration variation. When the Soret coefficient is negative, we have that $\delta F^{th}$, or $\Delta\Pi^{th}$, is positive for the solvent molecules, which will therefore concentrate at the cold side, while the solute will concentrate at the hot side due to Newton's third law. In general, saying that $\delta F^{th}$, or $\Delta\Pi^{th}$, is positive (negative) means in the DML that the radiation pressure due to the *lattice particle* stream on species "1" is higher (lesser) than on the species "2". Because the system is closed, it will obviously happen that only one of the two species will be pushed towards the cold side by the stream of *lattice particles*, while the other will move to the hot side due to Newton's third law.

Equations like (45) or (46) for $S_T$, or (44) for $D_p^{th}$, have the intrinsic issue that makes them not easily calculable. In order to get an expression for $S_T$ easy to be evaluated on experimental or numerical basis, we write eq.(52) at equilibrium replacing the expression for the thermal flux using the Fourier law; we get then:

$$53.\ \delta F^{th} = \sigma_p \cdot \Delta\Pi^{th} = \sigma_p \cdot \left[ \left( \frac{J_q^{wp}}{u_\phi} \right)_1 - \left( \frac{J_q^{wp}}{u_\phi} \right)_2 \right] = \sigma_p \cdot \left[ \left( \frac{K\, dT/dz}{u_\phi} \right)_2 - \left( \frac{K\, dT/dz}{u_\phi} \right)_1 \right].$$

At the Soret equilibrium, the power generated by the thermal engine is dissipated against the viscous forces:

$$54.\ W^{ph} \equiv W^{th} = W^{\eta};$$

where

$$55.\ \begin{cases} W^{th} = \delta F^{th} \cdot \langle \upsilon_p^{th} \rangle = \sigma_p \left[ \left( \frac{K\, dT/dz}{u_\phi} \right)_2 - \left( \frac{K\, dT/dz}{u_\phi} \right)_1 \right] \cdot \langle \upsilon_p^{th} \rangle \\ W^{\eta} = 6\pi\eta r_p \left( \langle \upsilon_p^{th} \rangle \right)^2 \end{cases} \Rightarrow$$

$$56.\ \sigma_p \left[ \left( \frac{K\, dT/dz}{u_\phi} \right)_2 - \left( \frac{K\, dT/dz}{u_\phi} \right)_1 \right] = 6\pi\eta r_p \langle \upsilon_p^{th} \rangle = 6\pi\eta r_p D_p^{th} \left. \frac{dT}{dz} \right|_{ext}$$

where the suffix *ext* stays for the external temperature gradient. It is trivial to get now:

$$57.\ D_p^{th} = \frac{\sigma_p \left[ \left( \frac{K\, dT/dz}{u_\phi} \right)_2 - \left( \frac{K\, dT/dz}{u_\phi} \right)_1 \right]}{6\pi\eta r_p \left. \frac{dT}{dz} \right|_{ext}}$$

Using again the Stokes-Einstein expression for the $D_p$ eventually we get for $S_T$:

58. $$S_T = \frac{\sigma_p\left[\left(\frac{K\,dT/dz}{u_\phi}\right)_2 - \left(\frac{K\,dT/dz}{u_\phi}\right)_1\right]}{K_B T_{av} \left.\frac{dT}{dz}\right|_{ext}} = \frac{\delta F^{th}}{K_B T_{av} \left.\frac{dT}{dz}\right|_{ext}} \cong \frac{\sigma_p\left[\left(\frac{K}{u_\phi}\right)_2 - \left(\frac{K}{u_\phi}\right)_1\right]}{K_B T_{av}}$$

Equations (45) and (58) provide for the first time a simple interpretation of the Soret effect at mesoscopic level, they are without matching parameters and can be verified by means of experiments. Equation (45) is based on elementary principles applied at mesoscopic level, while eq.(58) is based on macroscopic quantities. Both equations and their implications, as well as their capabilities in the comparison with experimental data of the Soret coefficient and of its sign, have been extensively analyzed in [18]. It is however mandatory to provide the direct comparison of such two expressions for $S_T$. Let's then consider firstly the third member of eq.(58); the numerator is the force exerted by the thermal field to displace the solute, or solvent, molecules so to generate and sustain the concentration gradient that gives origin to the Soret effect. At equilibrium the (average) energy available in the system is $K_B T_{av}$. The ratio between such two quantities is just the average displacement of the solute (solvent) *particle*, $\langle \Lambda \rangle$, due to the $\langle \delta \nu_p \rangle$ collisions with the phonons, $S_T = \frac{\delta F^{th}}{K_B T_{av} \left.\frac{dT}{dz}\right|_{ext}} \approx \frac{1}{\langle \Lambda \rangle \cdot dT/dz} \approx \frac{1}{\langle \delta T \rangle_z}$, that is exactly the same expression as eq.(46). Consequently, eqs. (45) and (58) represent two equivalent expressions for the Soret coefficient [18]. Experimental data available from the literature [see for instance 122,134,144-149,172-174] provide for $S_T$ values in the range $0.4 \cdot 10^{-3} °C^{-1} \le |S_T| \le 1.2 \cdot 10^{-3} °C^{-1}$ approximately, from which $\langle \delta T \rangle$ holds approximately $800°C \le |\delta T| \le 2500°C$. By supposing that such a temperature difference is extended over a few molecular diameters, the virtual temperature gradient, $\left\langle \frac{\delta T}{\delta z} \right\rangle \cong \frac{\Delta T}{\langle \Lambda_{wp} \rangle}$, would fall in the range $2 \cdot 10^3 \frac{°C}{nm} \le \left\langle \frac{\delta T}{\delta z} \right\rangle \le 6 \cdot 10^3 \frac{°C}{nm}$, approximately, confiming the assumption that the external gradient represents a small perturbation to the virtual gradient.

Before closing such section, we propose to the readers a speculative reasoning on the meaning and definition of $S_T$ in NET. In the phonon-*liquid particle* collisions of Figure 1a), part of the phonon energy is transformed into kinetic and potential energies of the *liquid particle*. Therefore, the interaction has an activation threshold for potential energy, and the amount of energy transferred to the kinetic energy reservoir of the *liquid particle* shall depend on how much energy is absorbed by internal DoF. When an external temperature gradient is applied to a system, the first effect that occurs is the establishment of the temperature gradient across the system, according to the Cattaneo-Fourier equation [24]. Therefore, once the internal DoF have reached the statistical equilibrium in terms of distribution of their degree of excitation depending on the local temperature, the phenomenon of matter diffusion takes place due to the temperature gradient and at the expenses of the kinetic energy reservoir. Inelastic effects dominate the dynamics during the transient phase, leaving the control to the kinetic reservoir at the steady-state, where only the elastic effects of the elementary interactions matter. Accepting the above reasoning, $S_T$ may be interpreted as the activation energy for the thermodiffusion process:

59. $$C(T) = C_0 \exp\left(-\frac{R/S_T}{RT}\right) = C_0 \exp\left(-\frac{1}{TS_T}\right) \approx C_0 \exp\left(-\frac{\langle \delta T \rangle_z}{T}\right)$$

where C is the solute concentration. Equations (59) means that if the energy $RT$ available at the temperature $T$ exceeds the activation energy $R/S_T$ , or alternatively, the average temperature of the system exceeds the temperature difference over $\langle \Lambda \rangle$ following the elementary interaction (see eqs.(45) and (46)), the thermodiffusoon takes place. It should be interesting to theoretically speculate on whether, and how, such activation energy is related to the presence in liquids of ***k**-gap*.

In [18], comparison with experimental data was provided for $S_T$, not only on its value, but also on its algebraic sign. It is important to highlight that the algebraic sign was found experimentally depending not only upon the couple of substances of a liquid mixture, i.e. solute and solvent, but also on the average temperature, thermal gradient, concentration, etc. The comparison revealed a good agreement with the experimental values within the order of magnitude, as well as the capability of providing the correct algebraic sign of $S_T$.

### *2.5 Momentum Transport and the "Unexpected" effect in liquids under shear in the DML.*

Modelling the viscosity in liquids is completely different from that in solids or gases, for several reasons, of which hereafter we report the main ones. Firstly, the viscosity of gases $\eta_g$ follows an increasing law depending exclusively on the temperature; in solids there is a linear relationship (Hooke-like) between the applied stress and the consequent deformation; in liquids it is the rate of deformation that is proportional to the applied stress, as given in eq.(61) below, representing the continuous counterpart to the linear relationship between the applied force and the speed of a material particle in a viscous medium. The viscosity of Newtonian liquids $\eta_l$ is experimentally found to exponentially decrease with $T$ , following the well known Arrhenius law $\eta_l(T) = Ae^{-\frac{b}{T}}$, where $A$ and $b$ are constants depending on the liquid nature and slowly dependent upon temperature [17,175-177]. Secondly, unlike in gases, forces working in liquids are intense as in solids, due to their similar densities. Thirdly, unlike in solids, liquid molecules not only oscillate around equilibrium positions, but also diffuse over distances comparable to, if not larger than, their size, similarly to the molecules of a gas. In short, they combine the two types of motions, of solid and gas, making very challenging the feasibility of a general theory of liquids [46]. The concepts shortly listed above were all taken into account by the several theorists who tried to frame the viscosity in liquids into a well defined physical model [178-191].

As in the previous sections, we start by adapting eq.(1) to the case of momentum transport[4]. Because such topic is typical of fluid-dynamic textbooks, it is not commonly found in thermodynamics treatises. However, by generalizing the concept of force-flux relationship given by eq.(1), one may simply introduce an equation expressing a linear proportionality between the momentum flux $J_p$, for instance along the *y*-direction, and the gradient along the *y*-direction of the momentum along the *x*-direction, $J_p(y) = -L_{pp}\nabla_y p(x)$, or more simply

$$60.\ J_p(y) = -L_{pp}\frac{d\upsilon_x}{dy}.$$

Here $L_{pp}$ is the Onsager coupling coefficient between the momentum flux and the shear gradient. Equantion (60) is the formal expression of a much famous equation, defining the liquid viscosity $\eta_l$ as the quantity connecting the shear force per unit area, $f_{x,y}$, to the shear gradient,

$$61.\ f_{x,y} = -\eta_l\frac{d\upsilon_x}{dy}.$$

Despite eq.(61) is known as the *Newton's law of viscosity*, he did *not* report it in his famous *Philosophiae Naturalis Principia Mathematica* [177]. Because in this manuascript we will deal only with the phononic contribution in liquids to the several physical quantities, we shall now contruct such contribution, $\eta_l^{wp}$, to the liquid viscosity. To do so, we return to eq.(14). Due to the symmetry of an isothermal medium, a similar equation can be written for the "$-z$" direction. Their summation will give a null diffusion over time because the number of scatterings along any direction is balanced by that in the opposite one. The algebraic summation of the "$+z$" and "$-z$" contributions gives the diffusion coefficient for the wave-packets $D^{wp}$

$$62.\ D^{wp} = \frac{\langle\Lambda_{wp}\rangle\langle u_{wp}\rangle}{3} = \frac{\langle\Lambda_{wp}\rangle^2/\langle\tau_{wp}\rangle}{3} = \frac{\langle\Lambda_{wp}\rangle^2\langle\nu_{wp}\rangle}{3}.$$

Let us take a liquid volume confined in between two walls, $d\upsilon_x$ being the relative velocity between the two walls. Therefore, a linear velocity gradient $\frac{d\upsilon_x}{dy}$ is applied to the liquid along the

[4] In [17] we adopted the opposite approach: we deduced firstly the expression for $\eta_l^{wp}$, the phononic contribution to $\eta_l$ in the DML, showing that it copes with the generic equation of NET coupling a flux with the correlated force

$y$-direction, normal to $x$, in stationary conditions. For each of the layers in which the liquid volume may be ideally subdivided, the frequency of elementary interactions of type a) and b) of Figure 7 along the $x$-direction shall be the same, so that their effect is null in the average along that direction. On the contrary, because of the presence of $\frac{d\upsilon_x}{dy}$, the frequency of elementary interactions is not the same for all the layers. If $\upsilon_1$ is the speed at which the layer '1' moves and $\upsilon_2$ that of the adjacent layer '2' , with $\upsilon_2 > \upsilon_1$, the viscosity of the liquid is the friction acting between two adjacent layers, from eq.(61). In the DML, the friction is mediated by the wave-packets. In Figure 7, a *liquid particle* A belonging to the Layer #2, moves at speed $\upsilon_2$ through the liquid in the $x$-direction, and a *liquid particle* B belonging to the Layer #1, moves at speed $\upsilon_1$, always in the $x$-direction. The *liquid particle* A is in an excited state and delivers an energetic wave-packet following the process b) of Figure 1; this crosses the border and hits the *liquid particle* B. As consequence of the emission, the *liquid particle* A goes in a de-excited state and slows down, due to the recoil. The opposite fate occurs to the *liquid particle* B: it acquires momentum and energy as consequence of the interaction with the wave-packet, as in an event a) of Figure 1. The acquired momentum and energy increase its velocity and the internal energy, exciting the collective DoF. The kinetic energy will be dissipated by means of collisions with the other particles of Layer #1, while the energy acquired by the internal DoF will be released into the thermal reservoir of Layer #1. This simple physical scattering mechanism allows momentum and energy be transferred from layer-to-layer by means of the wave-packets. In order to convert such idea into a formalism by means of the laws of physics, we start noting that the way in which liquid particles exchange momentum across the adjacent layers, as described above, copes with the definition of $J_p(y)$ in eq.(60), and is also consistent with the molecular picture of momentum transport and the kinetic theories of gases and liquids. This idea, so simple to seem naïve, all in all is the analogous treatment given for heat and mass transport in the DML, and copes with the wisdom that momentum flows from a region of high velocity to a region of low velocity. In agreement with eq.(60), the velocity gradient is assumed as the "driving force" for momentum transport, and the momentum flux as the "generalized flux", following the usual meaning of the NET. Let us then proceed with writing the equivalent of eq.(61) for the momentum flux carried by the *lattice particles*. The motion of the wave-packets driven by $\frac{d\upsilon_x}{dy}$ is an ordinary diffusive process for the wave-packets, therefore it is described by the diffusion coefficient $D^{wp}$ introduced in Eq.(62). The product of $D^{wp}$ times the density of *lattice particles*, $\rho_l^{lp}$, defines the dynamic viscosity $\eta_l^{wp}$, i.e. the momentum per unit of surface:

63. $\eta_l^{wp} = \rho_l^{lp} D^{wp} = \rho_l^{lp} \frac{\langle \Lambda_{wp} \rangle^2 \langle \nu_{wp} \rangle}{3}$

Any attentive reader shall have surely recognized that eq.(63) is gas-kinetic like, therefore in apparent contradiction with what stated above. However, such outfit is not surprising, but perhaps even awaited. In fact, $\eta_l^{wp}$ accounts only for that fraction, proportional to $m$, of the total viscosity due to the interactions between phonons and *liquid particles*. Therefore, both the inter-molecular interactions, and phenomena involving multiple *lattice particle* interactions are neglected. Said in other words, and as pointed out in [22,17], wave-packet$\leftrightarrow$*liquid particles* interactions are considered in the DML as gas-like interactions, also in view of the wave-packet concentration, see eq.(23). It is therefore not surprising that $\eta_l^{wp}$ is given by the product of $D^{wp}$ - that takes into account the momentum carried by the phonons – times the density of clusters, $\rho_l^{lp}$ – the material fraction of the liquid capable of exchanging momentum with phonons, accounting only for how many *liquid particles* are present per unit of volume, $n_{lp}$, and for their average mass, $m_{lp}$,

64. $\rho_l^{lp} = m_{lp} n_{lp}$.

Of course, it holds $\rho_l^{lp} < \rho_l$. Introducing the previous expressions into eq.(60), redefined for the wave-packet component, we eventually get

65. $J_p^{wp}(y) = -L_{pp}^{wp} \nabla_y p(x) = -\eta_l^{wp} \frac{d\upsilon_x}{dy} = -\rho_l^{lp} D^{wp} \frac{d\upsilon_x}{dy} = -m_{lp} n_{lp} \frac{\langle \Lambda_{wp} \rangle^2 \langle \nu_{wp} \rangle}{3} \frac{d\upsilon_x}{dy}$

Equation (65) defines the viscosity $\eta_l^{wp}$ following the usual meaning of the momentum transported per unit of surface by the wave-packets; its linearity ensures that we are dealing with Newtonian fluids. The readers interested to deep the development of such equation and of the other concerns beneath it, are warmly addressed to [17], where also its Arrhenious dependence is discussed, as well the comparison with experimental data, giving to it well funded arguments to assume $\eta_l^{wp}$ as the contribution to the liquid viscosity in Dual Liquid models.

Other than the comparison with the experimental data, a very robust test bench for the model of viscosity in the DML, that it is mandatory to report here, is provided by the first ever physical interpretation of the phenomenology discovered in liquids under shear strain by Noirez and co-workers [12,14,16,18], "unexpected" by the discoverers. It consists in the coupling between mechanical and thermal effects in isothermal liquids in shear motion, whose first theoretical interpretation was very recently provided in the framework of DML [17-18]. It was detected in two similar experiments, whose difference being only the motion law of a movable disk, as schematically represented in Figure 8. In both cases, one observes the occurrence of a temperature gradient in a liquid under shear, opposite to the shear gradient, due to the momentum transferred to the liquid by the moving plate. The phenomenon was believed classically unexpected for several reasons. First, following a purely fluid dynamical approach, the transformation of part of the work due to the external force, needed to overcome the liquid viscosity, into thermal energy is not admissible. At the speeds typical of the experimental setup, the stored energy cannot generate viscous heating according to the empirical evaluation based on the Nahme (or Brinkman) number, speeds close to that of sound being needed. Second, even assuming that a heating of the liquid is possible due to the internal friction, one would expect that the momentum transfer from the moving plate would induce a heating of the liquid layers facing the moving plate itself. On the contrary, exactly the opposite is observed [12,14,16], the heating of fluid being concentrated in proximity of the surface at rest. Because in the framework of the DML the effect is interpreted as a consequence of the *liquid particle* ↔ wave packet collisions, the above limitations are removed. Following the interaction, the energy moves from the fast layer to the slow one, as in Figure 9. Therefore, a current of wave-packets continuously crosses the medium, from the layer at high speed towards the layer at low speed. Figure 10 shows this behaviour in pictorial way. This induces a heating of the slow layers, in particular of that part of each layer adjacent to the slower one, and not of the layer adjacent to the mobile plate. In the DML, in fact, one expects exactly the onset of a thermal gradient, oriented in the opposite direction to the velocity gradient, i.e. from the faster layers toward the slower ones, due to the current of wave-packets, carrying momentum and energy, as experimentally detected. They are pushed by the external shear force from the fast layer to the slow one, where they accumulate. In the NET framework, this phenomenon is the response of the system to the external momentum gradient; it may be thought as similar to the thermodiffusion, discussed in section 2.4, where the phonons are pushed by a momentum gradient rather than by a thermal gradient (we will return on this point at the end of this section). The DML provides the first-ever physical interpretation of the phenomenon discovered by Noirez in a quantifiable form, as it is shown now. The velocity gradient $\frac{d\upsilon_x}{dy}$ represents the generalized force responsible for the momentum flux $J_p^{wp}(y)$ given by Eq.(65), to which the liquid reacts with a faint thermal gradient in the opposite direction. Other than the momentum flux, $J_p^{wp}(y)$ represents also the density of thermal energy due to the wave-packet' current. Therefore, dividing $J_p^{wp}(y)$ by the heat capacity, one gets the $\Delta T$ generated by the wave-packet current:

66. $$\Delta T = \frac{\overline{J_p^{wp}(y)}}{(\rho C)_l} = \frac{W_{wp}}{(\rho C)_l} = \eta_l^{wp} \frac{\rho_l^{lp}}{(\rho C)_l} \frac{\overline{d\upsilon_x}}{dy}.$$

Because the heating affects the whole liquid, in eq.(66) the "total" heat capacity of the liquid is used, while the vector modules $\overline{J_p^{wp}(y)}$ and $\overline{d\upsilon_x}$ are introduced to avoid sign confusion. The physical process progresses until the velocity gradient $\frac{d\upsilon_x}{dy}$ is working, the viscosity representing the manifestation of the tendency of any system to reach the equilibrium. Analogously to the case of the Soret coefficient, in [18] a different approach was used to calculate the temperature difference given in eq.(66). The equality of the two expressions is a confirmation of the rightness of the approaches, both based on the liquid duality. $\Delta T$ in eq.(66) can be estimated by means of the data contained in [18], so that one gets $\Delta T \approx 0.06K$, of the correct order of magnitude as measured and reported *ibidem*. This phenomenology, at the best of the author' knowledge, has never been evidenced in liquids (but also never looked for!, except unintentionally by the Noirez' group). In conclusion, it is reasonable to assume that even this aspect of the physical mechanism proposed for the viscosity in flowing liquids due to their duality, is explained by the DML. The onset of a temperature gradient opposite to the velocity gradient in the liquid is therefore not unexpected at all, but rather a crossed effect, typical in NET. In the DML, the viscosity is nothing else than a process that tends to balance the *liquid particles* and wave-packets populations with different momentum contents.

We want now to attract the reader' attention to two final considerations. Neglecting possible dissipation within the system, the final product of the shear flux injected into the system is a temperature difference $\Delta T$ due to the phonon current pushed by the momentum flux, $J_p^{wp}(y)$. In turn, $J_p^{wp}(y)$ is also the thermal energy flux due to the phonon current, $J_p^{wp}(y) = -\eta_l^{wp}\nabla_y \upsilon_x = \delta q_T = (\rho C)_l \Delta T$. If $\delta q_T$ is due to the imbalance of the phonon distribution throughout the liquid, as represented in Figure 10, it is given by

67. $\delta q_T^{wp} = \delta(n\varepsilon)^{wp} \cong \varepsilon^{wp}\delta n^{wp}$.

Ultimately, we have all the quantities available to calculate the phonon density step responsible for the temperature drop detected:

68. $\delta q_T^{wp} = J_p^{wp}(y) = -\eta_l^{wp}\nabla_y \upsilon_x = \varepsilon^{wp}\delta n^{wp}$ ➔

69. $\delta n^{wp} = \frac{\rho C_V}{\varepsilon^{wp}}\Delta T = \frac{J_p^{wp}(y)}{\varepsilon^{wp}}$.

Using the values given above for $\Delta T$, assuming for simplicity $\rho C_V \approx 1 cal/cm^3$ and $\nu^{wp} \approx 10^{12} Hz$ as deduced in [22], one easily gets $\delta n^{wp} \approx 4 \cdot 10^{20} cm^{-3}$, amounting to about $1\%$ of the value of the wave packet density estimated in an isothermal liquid at room temperature in eq.(23).

We finally propose here to the reader a final consideration on the crossed effect discussed in this section, namely a momentum flux in liquids generating a parallel thermal gradient. By looking at such coupled effect with the eye of the NET, one should expect that a similar crossed effect, although with the roles of momentum flux and thermal gradient exchanged among them, should exist in liquids, namely a temperature gradient in a liquid giving rise to a momentum flux. This is exactly the thermodiffusion, discussed in section 2.4, where the solute (or solvent) molecules are pushed by a momentum flux generated by a temperature gradient imposed to the system. Being the roles exchanged between the *liquid particles* and the *lattice particles*, we may consider the phenomenon discussed in this paragraph as the crossed effect to the thermodiffusion, with the *lattice particles* pushed by the momentum flux instead of the *liquid particles* pushed by a heat flux. Their accumulation on a side of the system gives rise to the faint thermal gradient experimentally observed. This explains also the agreement about the relevance of the effect of external perturbations with respect to what is the "normal" situation in a liquid at equilibrium, i.e. that either the influence of an external heat flux or of a momentum flux has on the virtual thermal gradient or on the phonon density.

## *2.6 The role of Instantaneous Normal Modes in the Dual Model*

The Phonon Theory of Liquid Thermodynamics, PLT, was proposed by a group of theorists a few years ago, based on a top-down approach. Among its main achievements, PLT provided for the first time an expression for the isochoric specific heat $C_V$ for liquids valid also for the solid, glassy, gas and quantum liquid states of matter, and confirmed by comparison with experimental data in 21 different liquids [35-38,41-45]. A very intriguing comparison between the results of PLT and those of DML was proposed in [23], where the readers interested to the full treatment are addressed; here only the main results shall be recalled, with particular attention to their comparison with the recent results of other authors [91,192-195] about the variation with temperature of the instantaneous normal modes, INM, discussed by Baggioli and Zaccone [52,54].

In the PLT, a modified form of the Hamiltonian is proposed, to account for elastic, or harmonic, $C_V^H$, and inelastic, or anharmonic, $C_V^A$, interactions. Ultimately, we may assume that in a liquid the total specific heat $C_V$ is contributed as:

70. $C_V = C_V^M + C_V^H + C_V^A = C_V^M + C_V^{wp} = C_V^M + C_V^H\left(1+\alpha T\right)$,

where with $C_V^M$ we have indicated the classical contribution due to molecular interactions, and $\alpha$ is the coefficient of isobaric thermal expansion of the medium. Of course, it also holds $C^{wp} = C_V^H + C_V^A$. With these assumptions, one gets the following limits for the parameter $m$:

71. $\frac{C_V^H}{C_V}(1+\alpha T) < m < (1+\alpha T)$,

We discover that $m$ and $\alpha$ are connected to each other. In the general case in which $\alpha > 0$, Eq.(71) becomes:

72. $\frac{C_V^H}{C_V}(1+\alpha T) < m < 1, \ \alpha > 0$

while the lower limit goes to zero with $C_V^H$. This conclusion is in line with the hypothesis of the DML. In fact, at high temperature icebergs melt and disappear from the liquid, and with them also the collective DoF contributing to $C_V^H$.

Remembering that it holds $\frac{dm}{dT} < 0$, an interesting expression can be deduced about the upper limit for $\alpha$, namely:

73. $\alpha < \frac{1}{T}\left(m\frac{C_V}{C_V^H} - 1\right); \ \frac{dm}{dT} < 0$

Because usually it is also $\alpha > 0$, we get the following relation:

74. $C_V^H < mC_V$

which represents an upper limit for $C_V^H$, providing the maximum thermal energy that can be stored into harmonic DoF of icebergs. Incidentally, this is in line also with the limits imposed to the total specific heat $C_V^{wp}$ [23]. Compiling the above expressions, we get

75. $\alpha > \frac{m^*}{m^2}\frac{dm}{dT}$,

as a lower limit for $\alpha$. We observe that the second member is a negative quantity; let us consider also those very few cases in nature where $\alpha < 0$, as for instance the water in the temperature range $(0 < T < 4)°C$ at atmospheric pressure. The minimum value identified for $\alpha$ in eq.(75) is valid of course also for $\alpha < 0$, and depends on the number of collective excitations in the liquid. Therefore, the complete range holding for $\alpha$ is:

76. $\frac{m^*}{m^2}\frac{dm}{dT} < \alpha < \frac{1}{T}\left(m\frac{C_V}{C_V^H} - 1\right)$.

By definition, $m$ accounts for the collective DoF actually participating to the energy distribution within a liquid. In a solid, the increase of $C_V$ upon temperature ($C_V \propto T^3$ Debye law) is due to the fact that the number of DoF which becomes excited, increases with T. In the DML, the liquid mesoscopic structure is a mix of solid clusters and amorphous phase. The fact that the number and size of clusters decreases with temperature is reflected by the temperature dependence of $m$ that, in turn, provides a decrease of $C_V$ upon temperature.

The calculation of $C_V$ in liquids is a debated problem of statistical physics and thermodynamics, often used as verification test of models of liquid state [116]. PLT and DML provides two independenly approaches to calculate $C_V$: on the one side, PLT provides the exact value of $C_V$ for the several states of matter; on the other side, the DML provides an expression based on the microscopic dynamics, where the only parameter still missing a direct evaluation is $m$. This parameter is linked to the so-called Vibration Density Of States, VDOS [50-51]. Besides, the DML identifies univocally both the harmonic and anharmonic contributions to the energy reservoir in liquids. Therefore, to correctly evaluate $m$ it would be necessary to evaluate the actual contribution of these two branches. A tool which could be of help in this context is that of the Instantaneous Normal Modes approach, INM, early considered by Zwanzig [32] in 1967 and by Stratt [54] in 1995 and recently fruitfully exploited by many authors. A generic phonon picture would require that the intermolecular forces can be regarded as harmonic, and liquids can simply not be viewed as being held together by springs. However, DML goes around this barrier with its duality based on the results obtained from numerical sumulations and recent experiments that have shed light on the actual arrangements of liquids at mesoscopic scale. A similar approach was however even adopted by other theorists. If used *cum grano salis*, the INM approach can either provide an accurate picture of the short-time and short-distances dynamics of liquids at a mesoscopic level, or, actually get insight on some of their properties from the statistical mechanics of liquids. Suppose we wonder how the potential energy of a liquid differs at some time $t_1$ from what it was at time $t_0$, with $t_1 - t_0 = \delta t$. What difference there is, it arises from the change in the liquid configuration, $\delta R = R_1 - R_0$. So, if $\delta t$ is short enough, and from this the "instantaneity", and $\delta R$ is small enough, we can expand the potential energy difference in powers of $\delta R$. Every new liquid configuration (that is, every new choice of time $t_0$) will have its own force and dynamical matrix, but for each choice there is a variable transformation that can be diagonalized, turning the cartesian coordinates of the atoms into collective coordinates, and our expression for the potential energy difference into a sum of independent harmonic contributions, i.e the eigenfunctions of the equations obtained by the Hessian matrix. This last sentence clarifies the attribute "normal" for such modes. What one calls the instantaneous-normal mode perspective, then, is that the dynamics of a liquid at short times is governed by precisely the dynamics of these modes, and that this dynamics, in turn, is nothing else than simple harmonic motion for the coordinates and the corresponding velocities. In this treatment the time evolution of a liquid configuration is completely prescribed by the set of instantaneous forces, frequencies and velocities. Thus, all that one needs to know are the probability distributions of these equilibrium quantities (as determined by the equilibrium distribution of liquid configurations) in order to compute the short-time behavior of any desired time correlation function. The answers so derived reflect of the dynamics of a variety of collective, but mutually independent, harmonic modes. Zwanzig [32] discussed accurately not only how to get the eigenvalues, but also the frequency spectrum and lifetimes of the INM thus obtained, finding an exceptional agreement with those values obtained half a century later by means of experimental techniques, and that we have deduced in the DML with an independent approach [see Table 1 in 22]. To complicate a little bit the situation is the fact that the eigenvalues, corresponding to the square of the frequencies, can either be positive or negative. Modes with positive eigenvalues are stable modes, corresponding to harmonic oscillations. Modes with negative eigenvalues are unstable, and correspond to solutions that exponentially grow or decay with time. The total number of modes is equal to the total number of DoF of the system. The INM spectum of solids is characterized by the presence of only real frequencies. In the case of liquids, a region of imaginary frequencies appears, the number of which increases with the temperature. Much effort was dedicated by as many authors [50-52,54,192,195-199] to identify the physical origin of the spectrum of INM, in particular of the imaginary one. There is a general agreement on the fact that the real eigenvalues are associated with the internal energy of the liquid, not the same is true for the imaginary eigenvalues. They were generally attributed as “unstable normal modes”, associated with diffusion or, more general, with anharmonicity. Without entering into the debates, we just want to propose here the interpretation of the INM spectrum within the DML framework. Firstly, a general conclusion directly derived from the INM analysis is that the number of real values decreases with $T$, in favour of the unreal values. Second, being the real values associated with harmonic elastic waves, typical of solids, it is arguable that in liquids, ephemeral aggregates, icebergs, form upon melting, whose size and number decreases with $T$. This interpretation is in agreement with the DML and with its duality. Heat and generally elastic energy, travel across the solid-like icebergs in a way similar to what they would make in solids. Indirect confirmations of this conclusion, that is much less speculative than it could seem at first sight, are the experimental results found for the dynamics of propagation of thermal energy in liquids. In particular, we cite i) that of Ruocco et al [69], who measured the values of about 3200 m/s for the speed of elastic waves in water over distances up to few molecular diameters, followed by the cut-off at about 1500 m/s on larger distances; and ii) the interesting results of Kayanattil et al. [88], who explicitely interpreted their results with the presence of solid-like regions in liquids, whose lifetimes are of the order of picoseconds. The ground becomes slippery when trying to provide a physical interpretation of the imaginary solutions, those characterizing unstable modes. Following the Frenkel picture, actually introduced even earlier by Maxwell and subsequently adopted by many other physicists of the past century, particle dynamics in liquids are characterized by oscillatory motions at quasi equlibrium positions, allowing liquids to show a solid-like behaviour, and by diffusive jumps, associated with viscosity, into neighbouring locations and enabling liquids to flow. Viscosity is both a microscopic and a dissipative phenomenon. One of the problems of statistical mechanics is precisely how such a dissipative transport mechanism arises from reversible microscopic motions of the medium's constituents, which in turn is related to the well-known Loschmidt paradox and the problem of the time's arrow. Following the physical interpretation provided by the DML of the viscosity, recalled in the previous section, momentum transport is due to the mechanisms shown in Figure 7, based on the presence of wave-packets. By nature, they are anharmonic, and precisely this aspect allows the momentum (and energy) be transferred to and from the icebergs upon collisions. Wave-packets are not present in (perfect) solids, being the energy tranported by means of harmonic waves. Upon melting, the icebergs form, and liquids reveal the capability of using wave-packets to allow icebergs to communicate among them, in order to guarantee the energy and momentum propagation anyway. Within the icebergs, energy propagates by means of (quasi)harmonic waves. As the temperature rises, given that the number and size of icebergs decreases, the number of stable, i.e. real eigenvalues, decreases, leaving the way to the unreal ones. Zhang et al. [198] have shown that the number of unstable modes follows an Arrhenious behaviour with temperature, as expected. The natural conclusion arrived at in the DML, therefore, is to associate the unstable normal modes with those characterizing the wave-packets interactions, by means of which they allow communication among the solid-like icebergs. They are long-lived collective excitations. Once again, the *lattice particle – liquid particle* elementary interaction, daughter of the dualism, plays the crucial role of mediating between the microscopic dynamics and the macroscopic behaviour of liquids.

## 3. The Time's Arrow paradox in the DML.

A dedicated paper is in preparation [113] where the entropy calculation for a dual system as the DML, especially in non-equilibrium configurations, will be analyzed in detail. Nevertheless, we want anticipate here that the DML, thanks to its duality, provides a possible physical interpretation of the apparent contradiction between the symmetry properties holding at microscopic scale and the deterministic time's arrow holding at that macroscopic. We will show that precisely the dualism of the DML, reflected by the the interaction between the two subsystems, is the key for removing and solving the paradox.

The so-called time's arrow paradox [96] is well known in thermodynamics and statistical physics since the time of Boltzmann, raised after the presentation of the H-theorem. It consists in a contradiction of the laws of physics, depending on whether they are applied on macroscopic or microscopic scale. In classical physics, time has a peculiar property that makes it different from any other variable: it always moves forward, never stops or goes backward. Sometimes physicists forget this property and fall into the trap of looking for physical interpretation of mathematical solutions which do not correspond to real-world. Newton's equations governing the dynamical evolution of a body under the influence of a force field give symmetric mathematical solutions with respect to time, regardless of whether they are applied to a single body or to a many-body system. The correct physical solutions corresponding to the real direction of the motion are identified by closing the Cauchy problem with the initial conditions of the real problem. In the four-dimensional world adapted by Einstein, after Lorentz and FitzGerald, in his formulation of relativity, the independence of time is seriously compromised, if not completely eliminated (although replaced with the speed of light). A different perspective is offered in classical physics by thermodynamics through the entropy state function, $S$ [5]. While the first law of thermodynamics simply adaptes the principle of energy conservation for many-body systems, the second law imposes specific limitations to the first law and explicitly assumes, throughout its formulation, that in nature not only time has its own direction at the macroscopic level, but also that this direction is clearly identifiable, as that in which $S$ increases with time in closed and isolated systems. It was exactly this autonomous vision of thermodynamics which revealed a paradox of the laws of physics, opening the possibility for physicists, as well as for philosophers, to speculate about their correctness [19,96-100]. This paradox consists in the evident contradiction between the time-symmetry property holding at microscopic level and the capability of thermodynamics of discriminating between past, present and future by means of the law of entropy increase. Boltzmann demonstrated the equivalence of his definition of entropy with that given by Clausius (for systems made by a large number of bodies) in the H-theorem, as well as the capability of being applied also to systems out of equilibrium, unlike the Gibbs entropy [97 and references therein, 200]. However, the debate, lasting since that time, is far from being closed and archived. The Boltzmann' mentor and colleague, Loschmidt [19], was the first to raise the paradox as a criticism to the H-theorem. He observed that, being the dynamics of the elementary constituents of a thermodynamic system temporally symmetric, it could not irreversibly evolve towards equilibrium. This argument was used by Loschmidt to deduce the second law of thermodynamics, describing the behavior of macroscopic systems, by demonstrating the contradiction with the time's arrow of (almost) all the elementary processes. This is called the "Loschmidt's Reversibility Objection". The reaction of Boltzmann to the first criticism was that the entropy increasing law just indicated the direction along with the time goes on. From this the origin of the term "time's arrow" conied by Eddington [96]. Later, Boltzmann reformulated the H-theorem and went around the paradox by ingenuously linking the entropy to the number of possible micro-states in the phase space in which the thermodynamic system may lie, $n$. Being such a number a very large quantity, he defined the entropy as proportional to its logarithm, $S = k_B \ln n$[6], so that $S$ can never be negative. From the practical point of view, the Boltzmann entropy reflects the fact that the number of micro-states available for an isolated system, i.e. the disorder of the system, increases vs time, as the dayly experience teaches. In this way, the macroscopic time evolution of a system becomes "by far the most probable" with respect to any other possible state. A common argument supporting this statement is that, following the classical mechanics, the universe is considered the only and surely most extended isolated system. However, it is equally true that it is far from being at the statistical equilibrium, the entropy decreasing, for instance, where celestial bodies, such as planets, stars or even galaxies, form and grew. Although such entropy anomalies, or discontinuities, are constrained in space and limited in time, representing only local perturbations for the whole system[7], the universe is still expanding. Definitively, the final destiny of entropy in classical mechanics is that of increasing. The situation is completely different if we assume that the dynamics on large scale is regulated by the general relativity instead of the Newton gravity law. In fact, dynamics of celestial bodies follows the space-time metric variation decribed by the metric tensor $g_{\mu\nu}$. When the dynamics of celestial bodies is discussed, the space-time metric properties are considered as "external conditions" to which they are subject. Therefore, not only the ensemble of celestial bodies cannot be considered as isolated, but also the external conditions are by no means steady in this case because $g_{\mu\nu}$ is a function depending on both space and time. The gravitational field cannot be included in a closed system and the whole universe itself is regarded as a system in a variable gravitational field. Consequently, the validity of the second law does not imply that the ststistical equilibrium is ever attained.

[5] The letter S is commonly used to indicate the entropy being the initial of the german word Störung, whose meaning is disorder.

In support of the deterministic character of the nature, clearly challenged by the Boltzmann statistical approach, Culverwell went back about 20 years later on the topic in a series of letters published on Nature [201-204]. In the first of them, after having exposed his strong criticisms to the "pulpubly absurd" propability option used by Boltzmann, he closes the letter with a much provoking question: "Will someone say exactly what the H-theorem proves?"

Zermelo's criticism [205] was based on the work of Henri Poincaré [206-207], an expert in the three-body problem, which, as is well known, has no exact analytical solution. Differently from the case of two-bodies, that can return on exactly the same path after a certain time, three bodies may only come arbitrarily close to an initial configuration, given "enough time". Poincaré established limits or bounds on the possible configurations of the three bodies from conservation laws. Zermelo argued that, after a long enough time, the particles would return to a distribution in phase space that would be indistinguishable from the original distribution. This is called the

[6] Actualy, the constant $k_B$ was introduced by Planck in 1900, who named it in Boltzmann' honor, in his manuscript on the black-body radiation.
[7] The question of whether the universe is closed or open, however, deserves a dedicated treatment, that goes outside the scope of the present manuscript.

“Poincaré Recurrence Time". Thus, he concluded, Boltzmann's formula for the entropy would at some future time go back down, contradicting Boltzmann's claim that the entropy always increases, as the second law requires. Interestingly, Boltzmann calculated the probability of a decrease of the entropy of a very small amount of gas of only a few hundred particles and found that the time needed to realize such a decrease was many orders of magnitude larger than the presumed age of the universe. Eddington was probably the first to deducing that the expanding universe provides a resolution to Zermelo's objection to Boltzmann.

Planck ingenuously changed completely the perspective being the first to realize that the cause of the entropy increase required by the second law was the interaction between matter and radiation, as finely illustrated by Doyle [208] (we will return shortly on the Boltzmann intuition). Einstein suggested in 1909 that the elementary process of radiation emission was irreversible [209-210]. A few years later, the advent of quantum mechanics brings back to the forefront the probabilistic option of the microscopic world with respect to the deterministic character of the macroscopic one.

After such short but necessary historical excursus, it is now time to face the time’s arrow paradox from the DML point of view. On the one side, we have observed that the elementary interaction between *liquid particles* and wave packets, Figure 1 and Figure 4, satisfies the postulate of microscopic time reversibility; therefore, it is compatible with the Newton, or Onsager, postulates. On the other side, it is also at the base of the thermodiffusion and even of the viscosity model, representing therefore the building block of dissipative macroscopic mechanisms. Consequently, it is also compatible with the Boltzmann statistical theory of time-asymmetric irreversible evolution towards equilibrium. The collective lattice excitations are the carriers of energy and momentum, allowing also mass transport, are linked to microscopic and macroscopic asymmetries, i.e. the virtual and external gradients, respectively, that give rise to directional currents of elastic excitation travelling across a medium. Our judgment of the time order in Figure 11 is not based on the dynamical laws of evolution alone, rather on the experience: one direction is common and easily recognizable, the other is never observed. *But why should this be so?* The Boltzmann idea to link the entropy definition to the availability of possible configurations explains in a natural way the observation, embodied in the second law of thermodynamics, that when a constraint is removed from an isolated macroscopic system, it evolves toward a state with greater entropy. To understand how this explanation works and at the same time ensure microscopic reversibility, let us analyze from the point of view of the DML what happens when the two bodies in Figure 11 are suddenly brought into contact. The number of DoF which excite increases by $2^N$, where $N$ is the number of particles, and the phase space of the wave-packets, carrying energy and momentum, increases too. This will continue until the complete system, A + B, reaches the equilibrium, corresponding to the highest possible value of entropy, or, that is the same, with the largest number of possible configurations for the system’ particles. After that time, we can expect to see only small fluctuations from equilibrium unless we wait for times that are much larger than the age of our universe and observe the two bodies resume their orginal temperatures. But what actually happens on the mesoscopic scale? Following Section 2.3, Figure 11 is the typical situation in which there is an imbalance in the number of interactions as those in Figure 1a. Therefore, we may replicate exactly the same reasoning. Due to the larger energy content of medium A with respect to B, the excess $\delta\langle \nu_p \rangle$ of interactions will diplace such an excess of thermal energy from the medium A to the B. This behaviour will proceed until the two media will have reached the same statistical temperature, so that the interactions of Figure 1a and Figure 1b will have the same probability of occurring. The situation is similar if consider two media with different mass concentrations, such as a pure liquid and a mixture of case b). In this case, by bringing the two media into contact, the concentration gradient shall generate a temperature gradient (Dufour effect) due to an excess of type 1b interactions with respect to those of type 1a, as described in Section 2.4. This *tug-of-war* between molecules and phonons, which develops on a mesoscopic scale by means of temporally reversible elementary interactions, manifests itself at the macroscopic level as the unidirectional flow of energy, or even of mass or momentum, depending on the real situation, univocally identifying the time's arrow. This was probably the mechanism Planck had in mind, and illustrated by Doyle, due to the interaction between matter and radiation, that in this case is the elastic, or thermal, energy carried by phonons. Figure 1 and Figure 4 together provide a mesoscopic solution for solving the paradox.

The paradox arose because both the laws of dynamics, regulating the microscopic elementary processes, and the second law, describing the evolution towards equilibrium of macroscopic multi-component systems, are experimentally verified. Looking closely to the paradox with the current knowledge, and particularly with the Planck's new perspective, leaves us understand that it arose because of the then-common belief that, even in non-equilibrium time-dependent conditions, collisions between molecules in a medium were believed to be random. That is, it was never thought that temperature, concentration or momentum gradient could trigger their preferential motion. On the contrary, this is precisely the innovative aspect of DML, due to the coexistence of the other subsystem constituting the liquids, the phonons.

Returning to Figure 11a, the redistribution of thermal energy is not the only result one gets once the two subsystems A and B are put in contact. In fact, the directive current of phonons will also push the liquid particles to displace from the medium A to the B, alike in a self-diffusion, following the dynamics shown in Figure 1a. This effect will generate, in turn, the retrodiffusion of either the liquid particles, or of phonons, following a sort of oscillating behaviour in the system. This will proceed until the residual energy shall be enough to sustain the energy current. These secondary effects are hardly detectable, and are even neglected in the formal treatment. Something of similar is true in the case of Figure 11b: once the diffusion starts, interactions like those of Figure 1b will give rise to a temperature gradient as the diffusion proceeds and the concentration gradient decreases. Once again, this process shall proceed until the residual energy will be enough to sustain it, and the effect will be increasingly weak and difficult to detect.

Although any attentive reader will certainly have already guessed, it is worth making an important, albeit banal, clarification. When a system is in equilibrium, that is, when there are no external perturbations that could trigger a preferential flow of phonons, and consequently of the energy, mass or momentum, the system "loses" not only the macroscopic but also the microscopic arrow of time [97-100], having the two interactions shown in Figure 1 the same probability of occurring and therefore, the same frequency of occurrence. They will be equally distributed, in the average, throughout the system.

A final note about the Poincaré recurrence time. The Poincaré recurrence theorem states that some dynamical systems will, after a very long albeit finite time, return to a state arbitrarily close to their initial state (for continuum systems). The Poincaré recurrence time is the time interval elapsed until the recurrence, whose length being generally extremely long, depending also on the degree of closeness. However, and importantly, it holds for isolated systems, called conservative systems. The theorem is commonly discussed in the context of ergodic theory, dynamical systems and statistical mechanics. The examples described above concern dissipative systems, so that the Poincaré theorem is not applicable. However, one may wonder whether it could, and should, be applied to both the subsystems constituting the liquid in the DML, the *liquid particles* and the *lattice particles*. The energy lost by the *liquid particles* is acquired by the *lattice particles* following the dissipative process. Consequently, the phase space of the former reduces, but what about that of the last? Answering this question imply the futher questions of how is the phase space of *lattice particles* defined and how to calculate it? These shall be hopefully answered in an future manuscript [113].

In conclusion, the DML, due to the presence of wave-packets, or *lattice particles*, which are thermal/elastic energy quanta, allows us to identify a time's arrow on a mesoscopic scale in condensed systems. Their interaction with molecular clusters, driven by virtual temperature or concentration gradients at equilibrium, or by the external gradients in systems out of equilibrium, identifies a privileged direction, a mesoscopic asymmetry, notwithstanding the elementary interactions remain time-reversal. It is therefore legitimate to wonder whether the macroscopic world is not really the sum of the microscopic worlds of which a system is composed.

Despite the many attacks faced by the second law, its primacy of unquestionability has held firm. Very famous are the words of Eddington in his book [96] "*If someone points out to you that your pet theory of the universe is in disagreement with Maxwell's equations, then so much the worse for Maxwell's equations. If it is found to be contradicted by observations, well, these experimentalists do bungle things sometimes. But if your theory is found to be against the second law of thermodynamics, I can give you no hope; there is nothing for it but to collapse in deepest* *humiliation*." Also famous is the sentence "*Nothing in life is certain except death, taxes and the second law of thermodynamics*" that Seth Lloyd wrote when argumenting about the probabilistic nature of the universe [211].

# 4. Summary, conclusions, and perspectives for theoretical developments and experimental verifications.

The DML is based on a simple yet profound observation: a liquid is neither a molten solid nor a condensed gas. It is both, depending on the time and spatial scale on which it is observed. The mesoscopic framework of DML overcomes the limitations of classical thermodynamics by treating liquids not as a homogeneous medium, but as hybrid system composed of dynamic "solid-like" structures (similar to continuously rearranging crystal lattices) immersed in an "ocean" of amorphous liquid. This model, funded on the dualism, effectively provides physical modelling of transport mechanisms, such as thermodiffusion and viscosity, and thermodynamic properties, such as heat conductivity, specific heat, Soret coefficient, diffusion coefficient, by integrating structured and disordered behaviors. By capturing the dual dynamic nature of liquids, the conceptual framework of the DML fits their nature into a modern landscape where no single model can describe all liquid properties. It is based on experimental evidences on the mesoscopic organization of liquids, and finds confirmation by comparison with experimental data of the transport parameters cited above. The keystone of the DML is the duality, that manifests through the interaction between the two populations constituting the liquid, the *liquid particles* and the *lattice particles*. The interaction has several peculiarities, namely: it is time-reversible, satisfying the Newtonian, and Onsagerian, dynamical requisites; it allows exchanging energy and momentum among the two populations, giving rise to the macroscopic phenomena observed in liquids, i.e. the transport of energy, mass and momentum, also in non stationary conditions, thanks to the tunnelling. Besides, it is also at the base of irreversible properties of liquids, such as the viscosity and the thermodiffusion. It represents the bridge between the microscopic world described by the statistical physics, and that macroscopic, described by the non-equilibrium thermodynamics. Finally, the interaction may explain the apparent contradiction between the microscopic reversibility and the macroscopic non-reversibility of transport phenomena in liquids. The most challenging aspect of this framework probably lies in the attempt to account mesoscopically for what classical thermodynamics freezes in static time averages: the competitive coexistence between high-frequency elasticity (solid-like phonon modes) and low-frequency fluidity (gas-like diffusive jumps). In fact, thermodynamics accounts only for physical quantities related to matter reservoir, not to that pertaining to the *lattice particles*. The definition and calculation also of entropy should be extended to both the interacting systems. To the best of the author' knowledge, this has never been done, and for this reason it will be the task of a future work [113]. The coexistence is approached from the perspective of statistical physics and non-equilibrium thermodynamics, by focusing on how the material component, the *liquid particles*, exchange energy and momentum with the *lattice particles*, through anharmonic wave packets, in order to determining their contribution to the physical parameters characterizing the liquid medium. Harmonic oscillations and anharmonic wave packets are the ways in which energy and momentum propagate within a liquid. Harmonic oscillations are supposed to propagate within the ephemeral icebergs, while *liquid particles* communicate among them and with the disordered liquid by means of anharmonic wave-packets, enabling the exchange of energy and momentum. The simplifying hypothesis that the collisions between wave packets and pseudo-crystalline structures have a single-particle character, as in the Boltzmann-Maxwell statistics, is supported by the wave packet' density estimation. The mesoscopic character of the DML allows investigating the intimate mechanism of interaction of *liquid particles* with lattice modes; in this way DML, providing the answer to the question of "why" liquid particles oscillate, indirectly answers also the question "whether" liquids are held together by springs; consequently, provides a way for calculating the order of magnitude of $\tau$ in ordinary liquids. The relaxation time or, what is equivalent, the characteristic time of the elementary springs, is one among the key-points of DML.

Yet, there are several open points, either theoretical or experimental, to be further investigated, such as for instance: the definition and calculation of the entropy and of its rate of production, the calculation of $m$, the experimental confirmation of the presence of the elastic perturbations induced by the application of external stimuli to liquids, just to cite few. The first point has been extensively discussed in previous sections. The calculation of $m$ is instead still an open point. It is worth noticing that $m$ is not a free parameter figuring *ad hoc* into the equations to tuning them to the expected results, but rather a quantity that can be quantified based on a statistical estimation of the DOF available for a given chemical species. In this field, a precursor method is that proposed by Ghandili [93].

On the experimental side, there is a lacking of data, because of the facts that the Dualism of liquids is a recent theoretical understanding, so that experimentalists have not specifically addressed their efforts on that topic. The "unexpected" behaviour of liquids under shear discovered by Noirez is a typical case of serendipity, and it was only recently theoretically framed by the DML. Accompanying this result are those achieved by Kayanattil [88] discussed above, while the results obtained between the end of past century and the beginning of the present, [67-87] were not clearly framed upon having been detected.

Apart from the above considerations, there is a large area of experimental configurations that was never investigated and therefore is still "*terra incognita*". This notwithstanding experiments carefully addressed could potentially provide very interesting results and shad light on whether the dual modelling of liquids is correct or not. The experimental configurations which we refer to is to investigating the modifications to which a liquid is subject during the transient state following the application of an external stimulus, i.e. exactly that part of an experiment normally avoided or skipped. In the DML in particular, the transient phase is of huge importance because it is exactly during it that the avalanche of wave-packets is oriented by the external gradient. Similarly, the correlation lengths increase affecting transport phenomena, such as viscosity, giving rise to potentially unpredictable phenomena. Capturing these manifestations from experiments would clear the field of hypotheses of any doubts regarding the mesoscopic structure of liquids. Possible experimental configurations could be designed. Starting from a liquid in isothermal, stable and controlled conditions, placed in a closed container with two opposite sides thermally controlled. Changing the temperatures on both sides, increasing that of the top side and lowering that of the bottom (to avoid convection), and having care of keeping as more as possible unchanged the temperature in the middle of the container, is a configuration suitable to analyse the dynamic evolution of the system, at mesoscopic scale, from the isothermal to non-isothermal states. An alternative way is to heat a small volume of liquid by hitting it with a focalised high power laser beam. A possible variation of this configuration is that of starting from a stabilized thermal gradient upward oriented (again, to avoid convection) and lowering the temperatures on both sides, up to reach the freezing point in the middle of the liquid volume. What will the evolution be of the liquid-to-solid transition? Will the liquid freeze at the same temperature? Will the local domains be oriented following the external temperature gradient, allowing in turn increasing the correlation lengths, sound velocity, thermal conductivity, etc., along the preferential direction of the external temperature gradient? Will the solid phase again isotropic or will it be influenced by the anisotropic thermal field? Understanding how liquid parameters, such as correlation lengths, sound speed, thermal conductivity, etc., evolve from equilibrium to non-equilibrium conditions is an interesting topic that deserves more attention from experimentalists, similarly to understanding whether and how a temperature gradient influences viscous coupling.

It is known that the experimental setup described above could be very difficult not only to be reached but also to be kept stable for time long enough to allow data to be collected. To avoid such issues, the best environment where performing such experiments is that of microgravity, i.e. that reached onboard platforms orbiting around earth (or even in a Lagrange point). This environment could be of huge importance also to investigate the intriguing theoretical findings of Esposito and Nicolis [212-213] on the negative gravitational mass carried out by phonons. In their papers Esposito and co-workers show that propagation of sound waves, constituted by phonons, through condensed media is associated with a negative gravitational mass. Indeed, in a density gradient they move towards a medium with a lower density, where the impedance is higher and their speed is lower. Therefore, the pivot of the reasoning is again the entropy of the system and how is it influenced by other factors, such as the gravity: a first as elementary as intriguing question is how the II Principle should be reformulated for a system in thermal equilibrium and immersed in a gravitational field, in which phonons subtract heat to the thermal reservoir upon collision with the liquid particle. Is there a “Demon” hounting the daily physics reality by making more reliable those events that are usually neglected? [214]. Why do phonons prefer the more difficult path than the easier one? What makes easier the path that is normally the most difficult?

The fact that the origin of that unlikely state of matter called "life" is linked to liquids does not surprise us that much, given the variety of their panorama [215].

# 5. Figures

## Figure 1

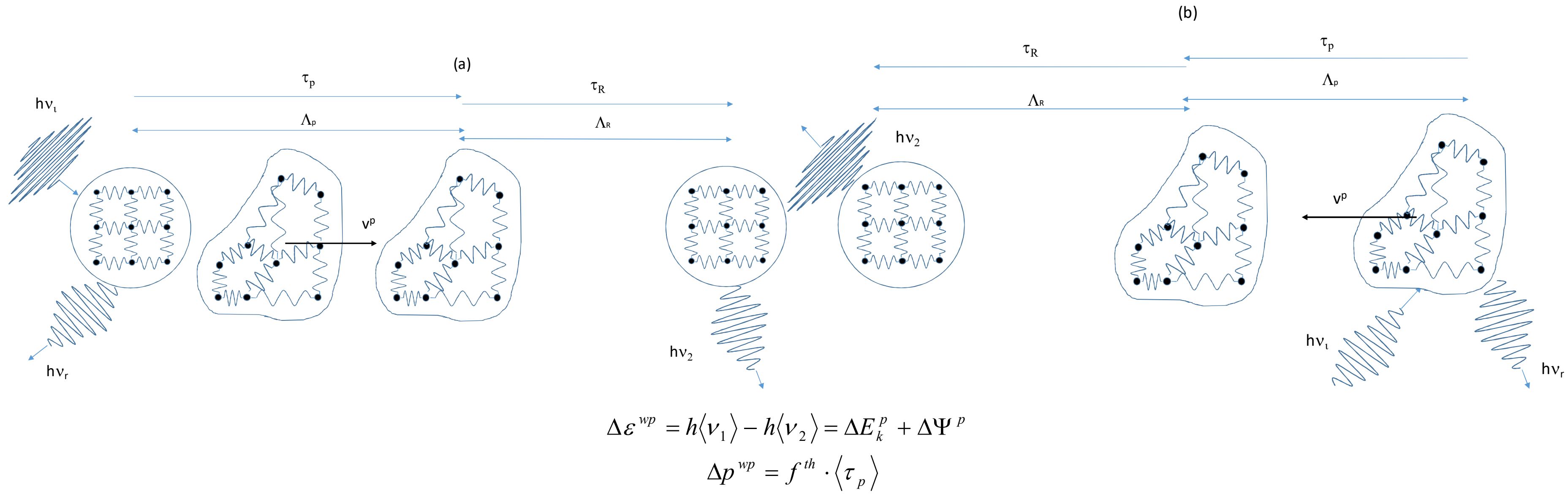

**Figure 1.** Schematic representation of inelastic collisions between wave-packets and *liquid particles*. This is the elementary interaction which the DML is based on.

# Figure 2

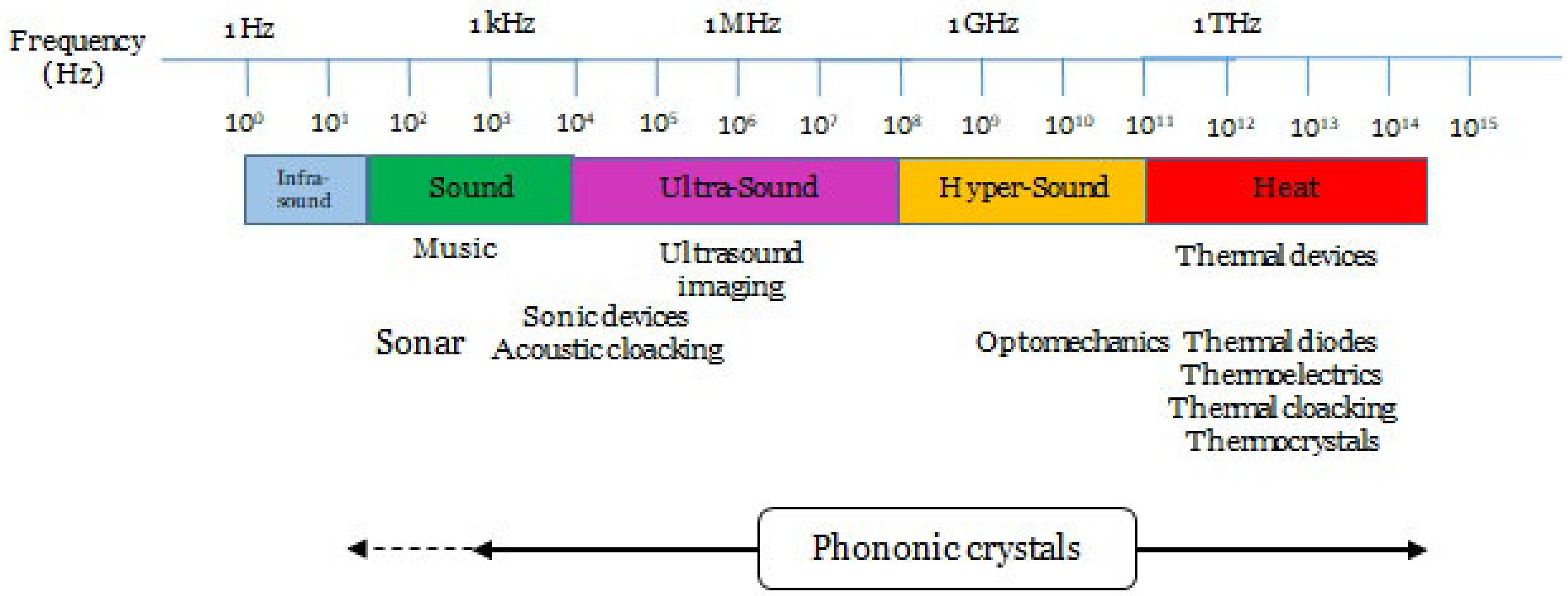

**Figure 2**. Frequency spectrum of the elastic perturbations propagating through media, with indication of the phononic spectrum.

**Figure 3**

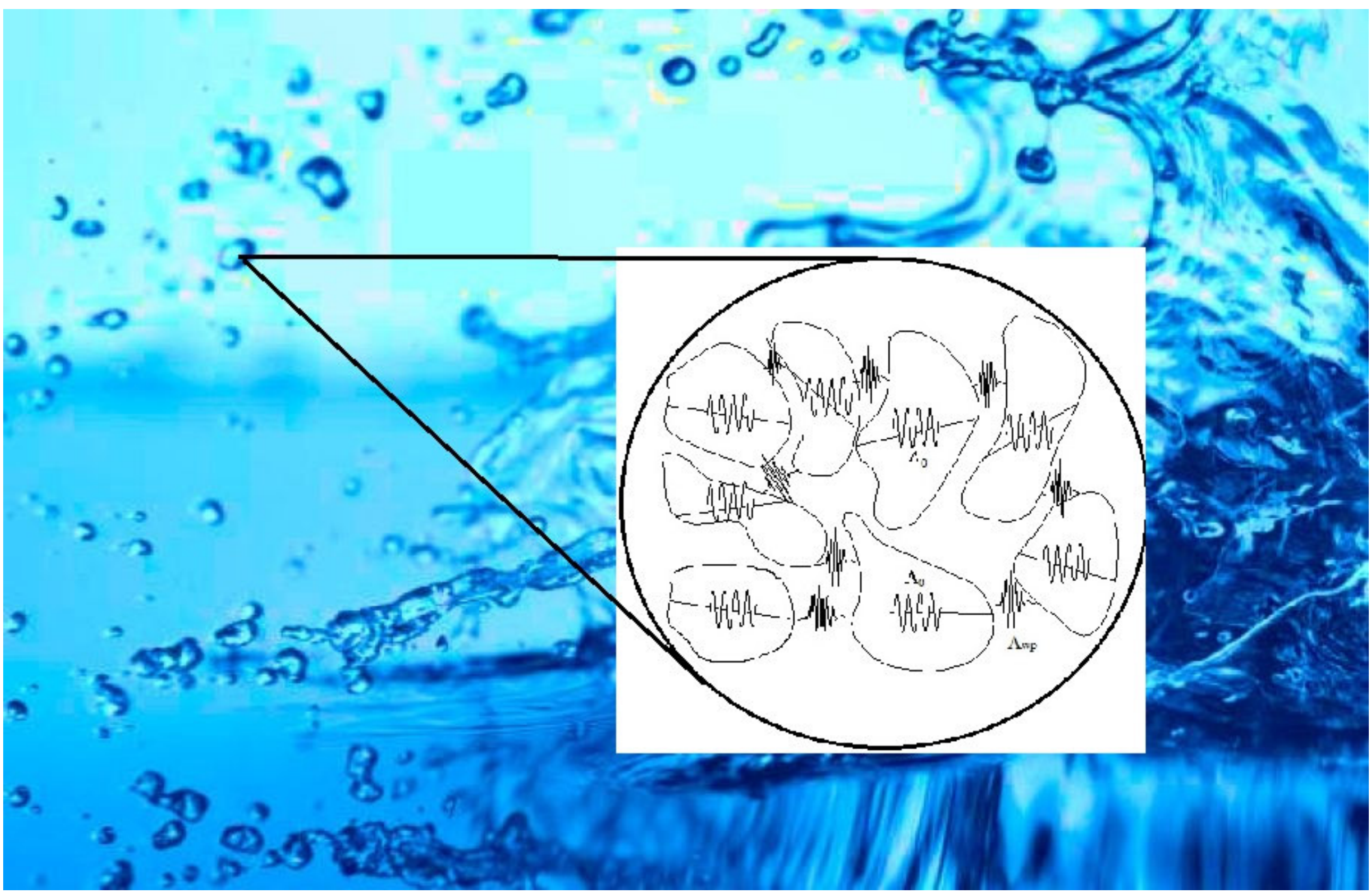

**Figure 3**. The pictorial model of liquids at mesoscopic scale, on which the DML is based: icebergs of solid lattice fluctuating and interacting within the liquid global system at equilibrium. It reflects also what may be deduced from experiments performed with IXS techniques [67-88], able to observe liquids at such scale-lengths.

## Figure 4

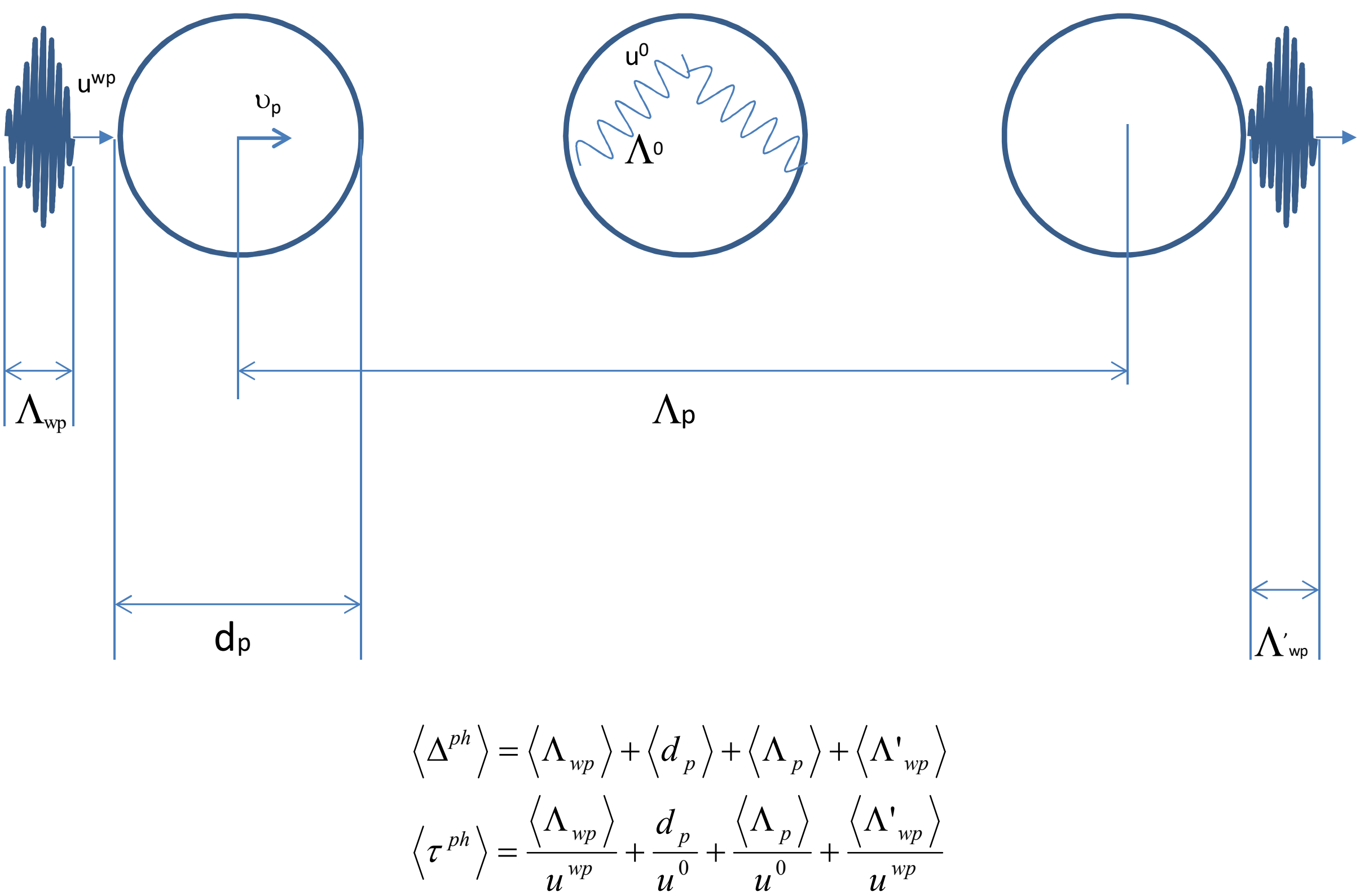

$$\langle \Delta^{ph} \rangle = \langle \Lambda_{wp} \rangle + \langle d_p \rangle + \langle \Lambda_p \rangle + \langle \Lambda'_{wp} \rangle$$

$$\langle \tau^{ph} \rangle = \frac{\langle \Lambda_{wp} \rangle}{u^{wp}} + \frac{d_p}{u^0} + \frac{\langle \Lambda_p \rangle}{u^0} + \frac{\langle \Lambda'_{wp} \rangle}{u^{wp}}$$

**Figure 4**. Close-up of the wave-packet – *liquid particle* interaction shown in Figure 1a. Only the first part of the interaction is represented, i.e. that during which the wave packet transfers momentum and energy to the liquid particle.

MDPI Credits

## Figure 5

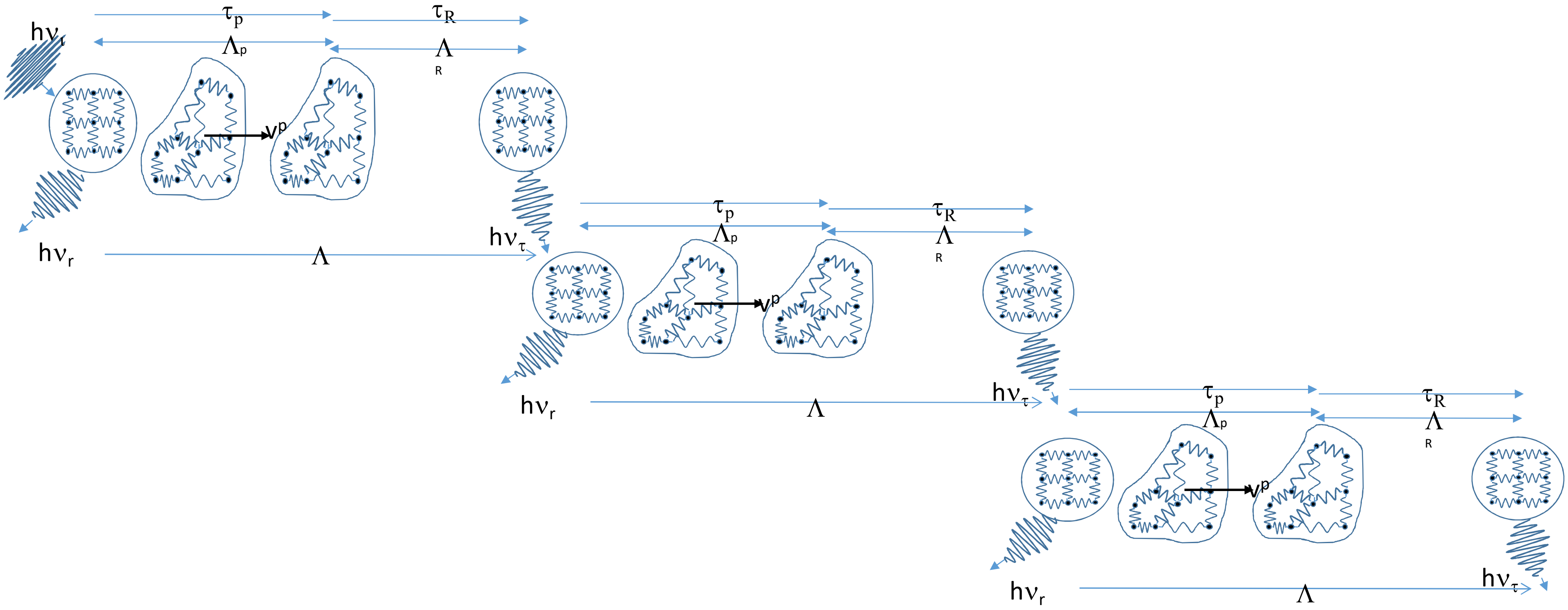

**Figure 5.** Avalanche process of successive wave-packtes – *liquid particles* interactions of Figure 1. Following the interaction, the force $f^{th}$ acts for $\langle \tau_p \rangle$ seconds displacing the particle by $\langle \Lambda_p \rangle$. Energy $\langle \Delta\varepsilon^{wp} \rangle$ and momentum $\langle \Delta p^{wp} \rangle$ are acquired by the *liquid particle* during $\langle \tau_p \rangle$, increasing the kinetic and potential energy of the last, and exciting its internal vibrational energy levels. The internal DoF oscillate similarly to those pertaining to solid state, giving tise to (quasi) elastic waves with characteristic wavelength $\langle \Lambda_0 \rangle$ and period $\langle \tau_0 \rangle$. $\langle \tau_R \rangle$ seconds later and $\langle \Lambda_R \rangle$ meters forward they are given back to the wave-packet reservoir, and the process is repeated again $\langle \tau_{wp} \rangle$ seconds later and $\langle \Lambda_{wp} \rangle$ meters forward.

## Figure 6

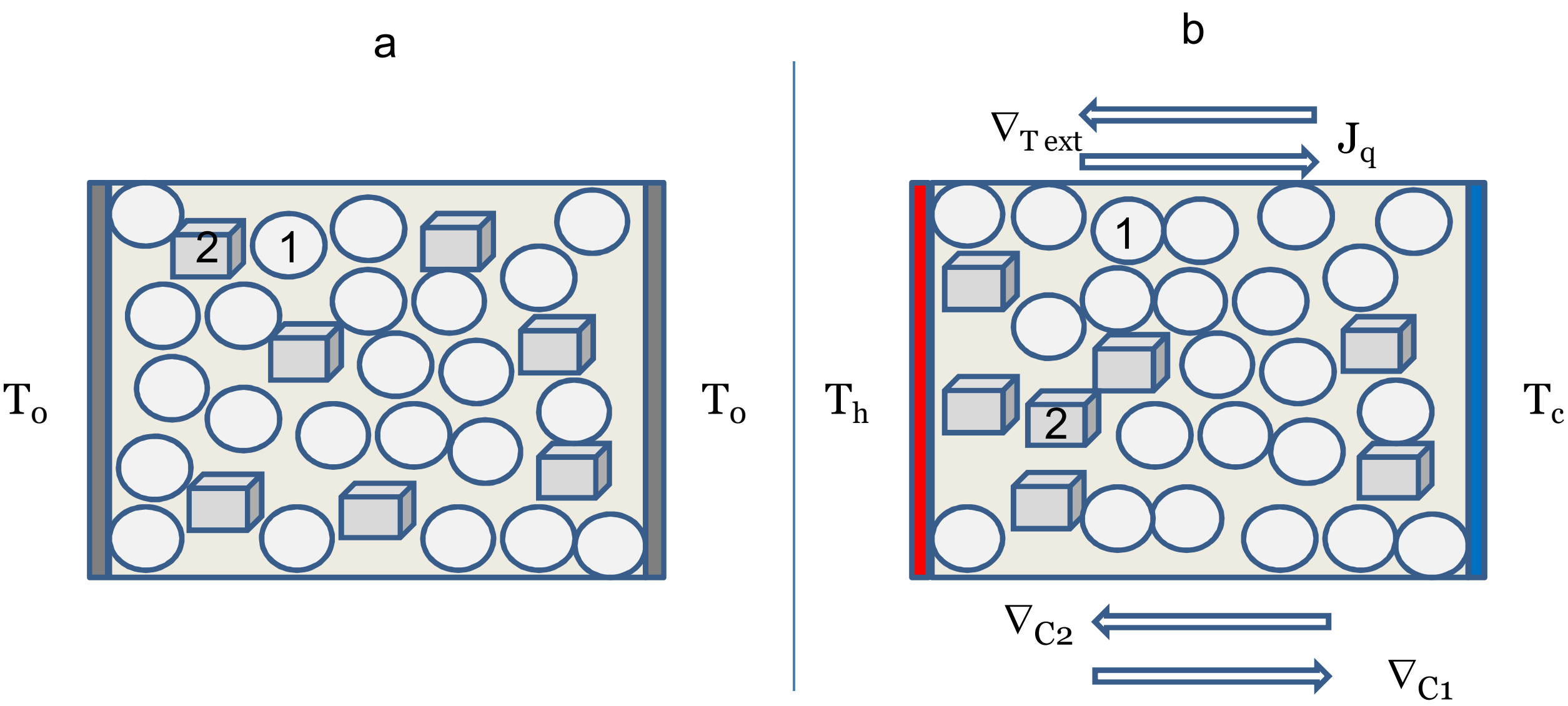

**Figure 6.** On the left side, (a), a mixture made by two different chemical species, 1 and 2, ideally represented by spheres and cubes, uniformly distributed. A temperature gradient $\nabla T_{ext}$ is applied to the medium generating a heat flux $J_q$. (a) represents the beginning of the experiment. As far as the heat crosses the medium, the two species separate, until a steady state is reached, characterized by the separation of the two chemical species, on the right side (b). The steady state is the Soret equilibrium, characterized by the dynamical equilibrium reached by the two diffusive mechanisms, one driven by the temperature gradient, the other by the concentration gradients of the chemical species.

**Figure 7**

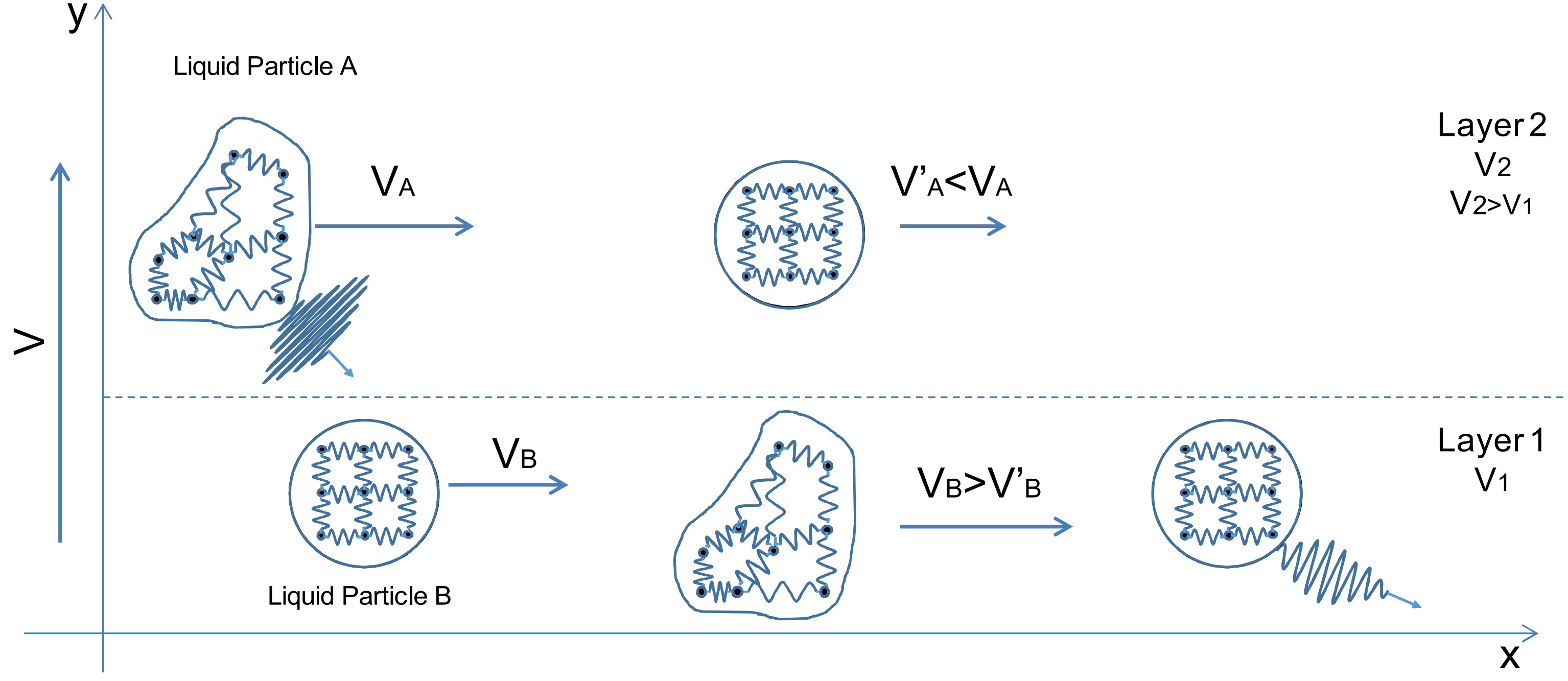

**Figure 7.** Schematic representation of the physical mechanism at the base of the viscosity in the DML, based on the elementary interaction shown in Figure 1 and Figure 4. $\upsilon_1$ is the speed of the Layer '1' and $\upsilon_2$ of the adjacent Layer '2', with $\upsilon_2 > \upsilon_1$, for simplicity both along the $x$-direction.. The viscosity is due to the friction between the two layers, the force between them working like a shear stress, as the fast layer pushes the slow one. In the DML, such friction is mediated by the presence of the wave-packets, responsible for the momentum (and energy) transport pertaining to the solid-like collective DoF. The *liquid particle* A belonging to the Layer #2 is in an excited state and delivers an energetic wave-packet following the process b) of ; this crosses the border layer and hits the *liquid particle* B. As consequence of the emission, the *liquid particle* A goes in a de-excited state and slows down, due to the recoil. The opposite fate occurs to the *liquid particle* B: it acquires momentum and energy as consequence of the interaction with the wave-packet, as in an event a) of Figure 1. The acquired momentum and energy increase its velocity and the internal energy reservoir, exciting the collective DoF. The kinetic energy will be dissipated by collisions with the other particles of Layer #1, while the energy acquired by the internal DoF will be released into the thermal reservoir of Layer #1**.**

**Figure 8**

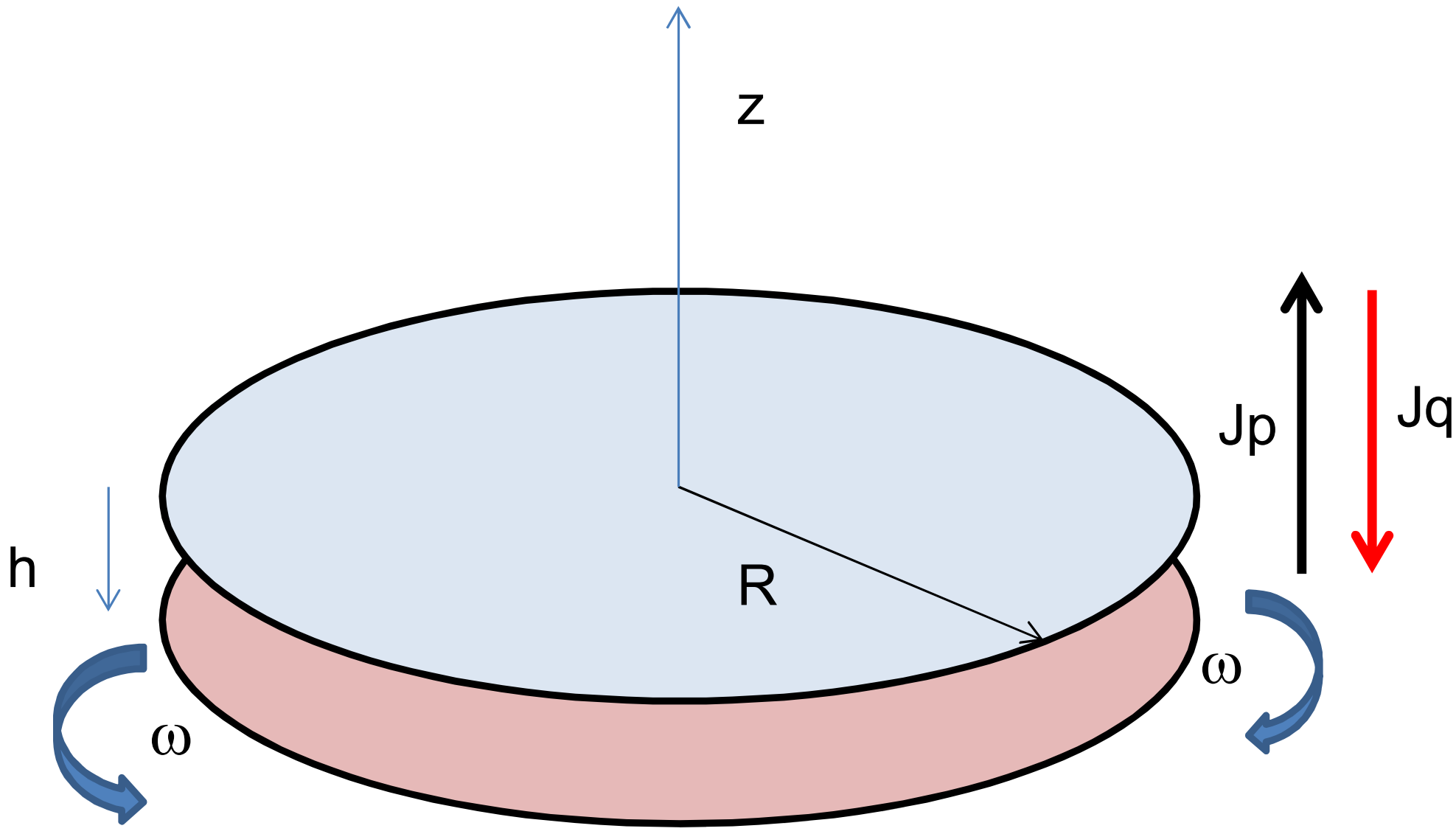

**Figure 8.** Schematic drawing (not at scale) of the experimental device adopted in [12]. The lower disc is fixed, the upper disc is driven by an external motor and drags the liquid, glycerol in this case, lying in between the two plates.

**Figure 9**

System at equilibrium

System out of equilibrium

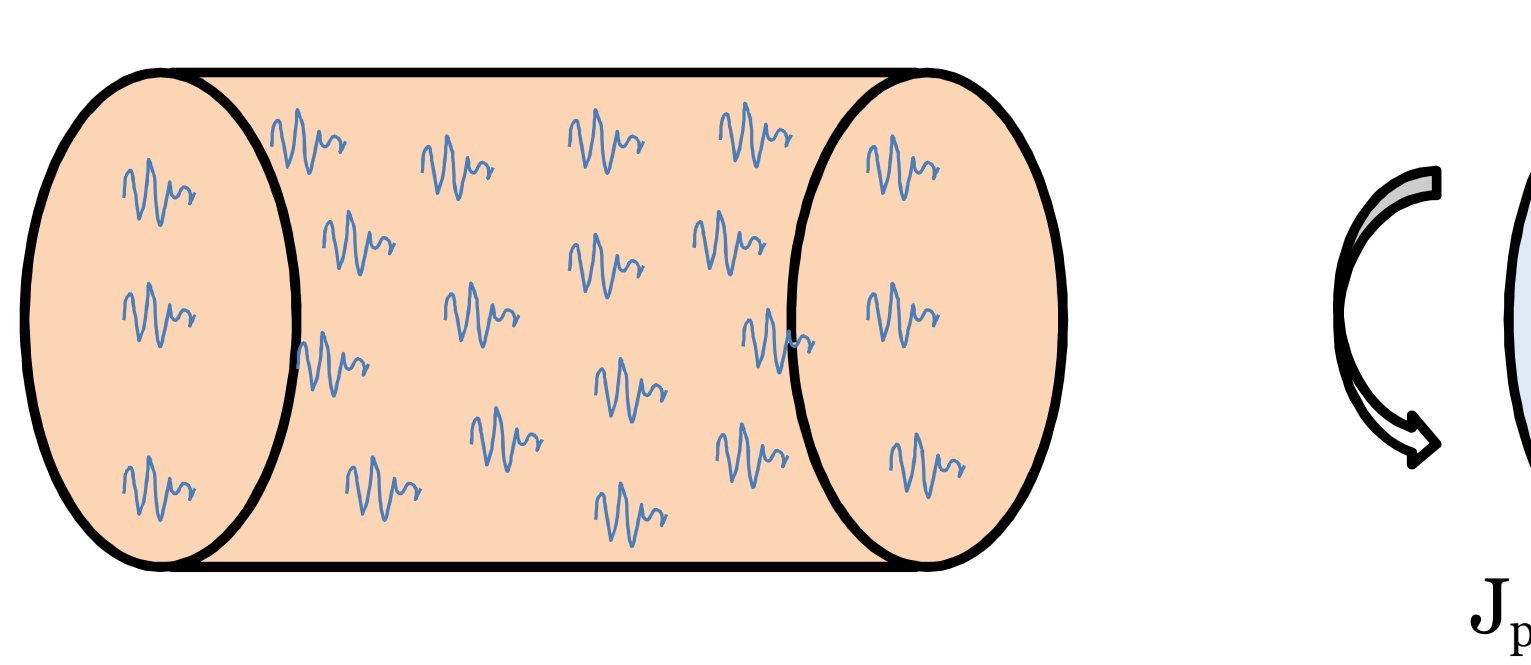

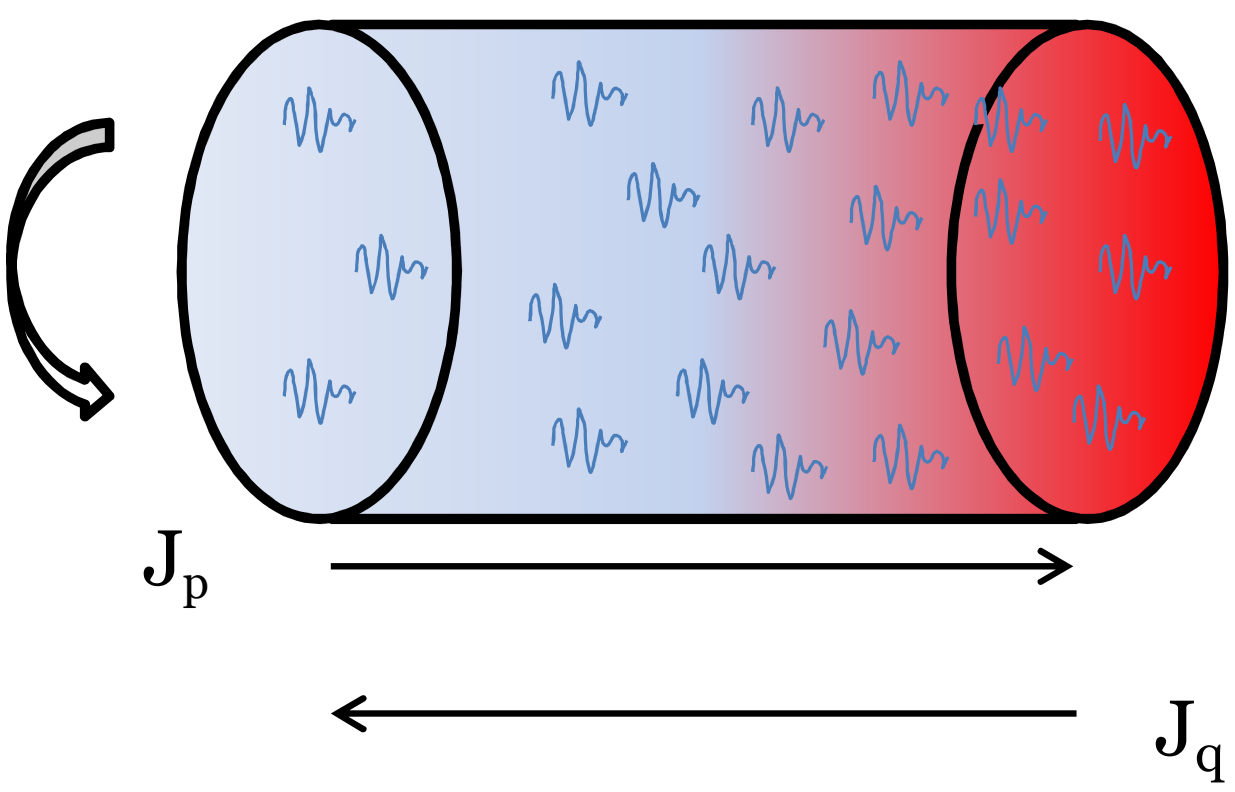

**Figure 9.** Drawing simulating the effect of the rotation of the upper plate of Figure 8 on the wave-packets, that are pushed towards the fixed surface, giving rise to a thermal gradient opposite to the velocity gradient.

**Figure 10**

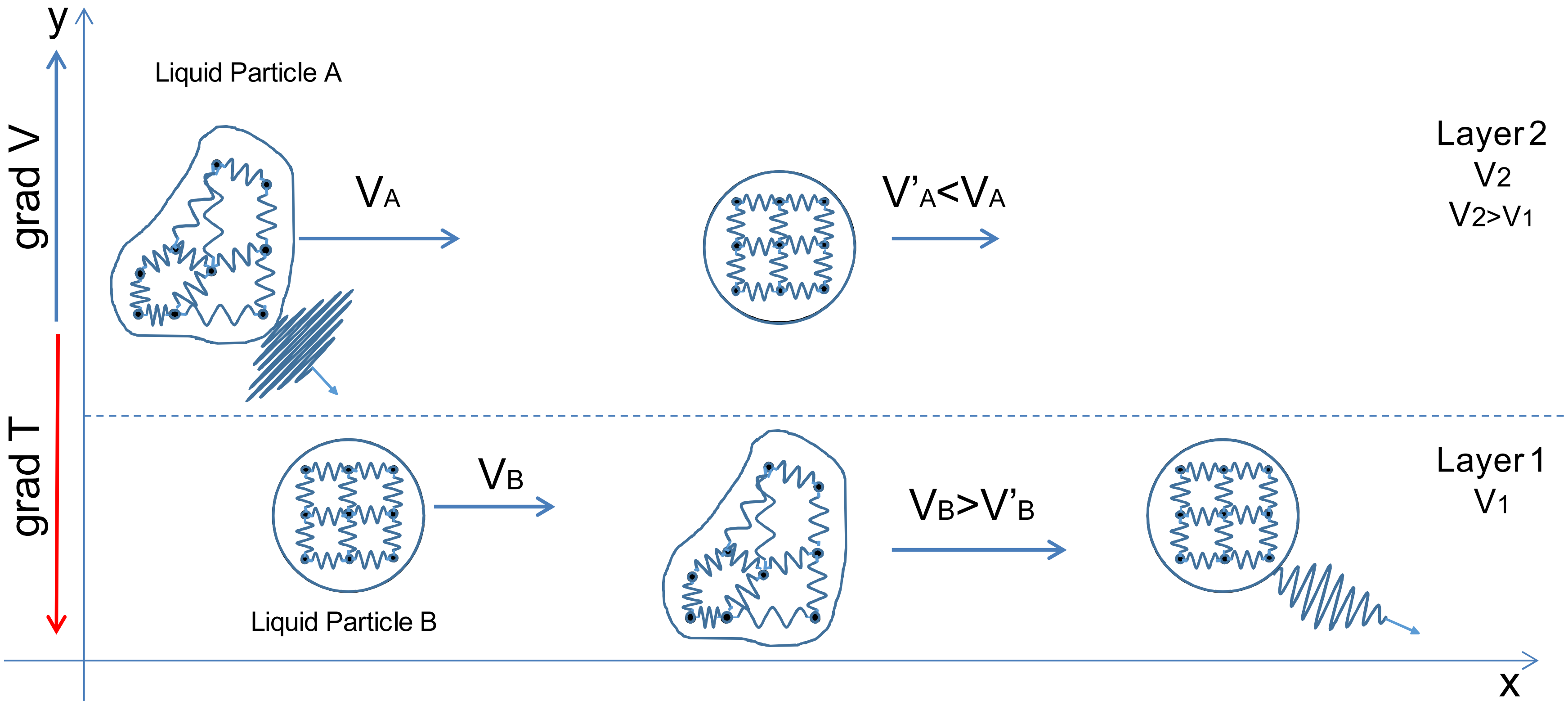

**Figure 10.** Same as Figure 7 but with indicated the external velocity gradient and the induced thermal gradient due to the current of wave-packets.

## Figure 11

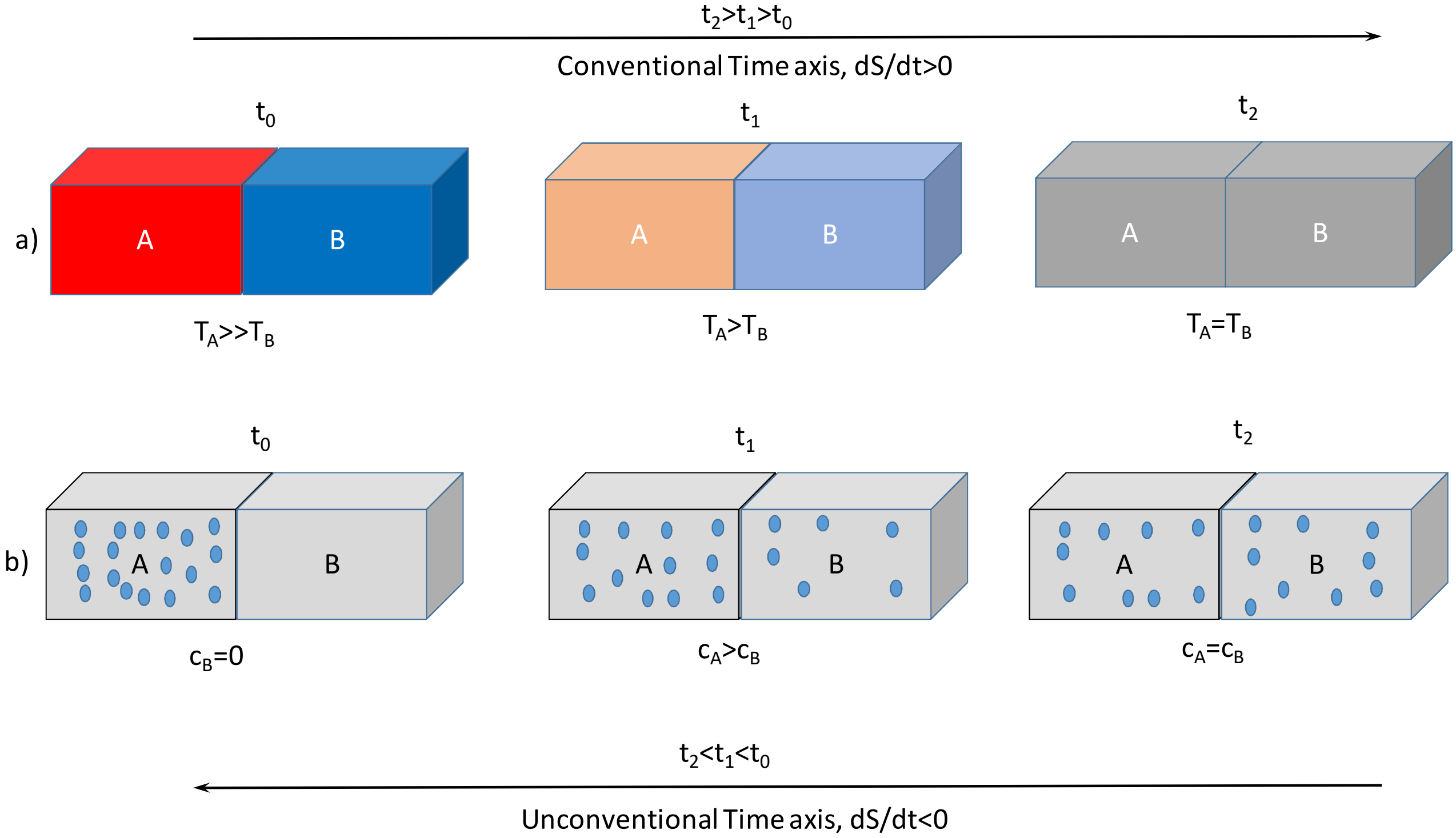

**Figure 11**. Imagine we have two bodies, A and B, with $T_A > T_B$ in a) and $c_B = 0$ in b), at $t = 0$. In a), the two bodies are separately in equilibrium. Let us introduce an external perturbation by placing the two bodies in contact with each other at $t = 0$ and film with a thermo-camera the "spontaneous" evolution of the temperatures of the two bodies. On the only basis of the dayly experience, that is nothing else than our unconscious perception of the laws of physics, anyone can state, without fear of contradiction, that will observe an increase in the temperature of B and a decrease in that of A, until the two bodies shall reach an equal temperature: the heat flows "naturally" and "spontaneously" from the body A, at higher temperature, to the B, at lower temperature. Let's now imagine projecting the film in reverse once the thermal equilibrium is attained; anyone watching this film without knowing that it is projected backwards, noticing that the temperature of A increases as well as that of B decreases "spontaneously" despite the two bodies being in contact, would argue that there is a trick involved, or a Maxwell' demon who is playing a joke. The situation is not so much different in b), where the starting conditions are of a medium A consisting in a mixture, and the B of a pure liquid. The two media are separated by a wall, that is suddenly removed at $t = 0$. As time goes on, the diffusion proceeds and the concentration decreases in A and grows in B, up to $t = t_2$, when the two media reach the same concentration. In the DML, the evolution of the two systems is due to the current of wave-packets, pushed by the temperature gradient in a), or by the concentration gradient in b). In the first case, there is a prevalence of collisions as those of Figure 1a, and of those of Figure 1b in the second case. These simple examples serve to highlight not only the relevant difference between the time evolution of macroscopic variables in non-equilibrium statistical thermodynamics compared to that of physical quantities, but also of the microscopic processes taking place within a complex system like those subject of statistical thermodynamics.